\pdfoutput=1
\documentclass[11pt,twoside,a4paper,cmspaper,final,collab]{cms-tdr}
\def\svnVersion{e4e2591}\def\svnDate{2025/09/03}\def\cmsCernNoTag{CERN-EP-2025-021}\def\cmsCernDate{\today}\def\cmsMessage{Published in the Journal of High Energy Physics as \href{http://dx.doi.org/10.1007/JHEP08(2025)156}{\doi{10.1007/JHEP08(2025)156}.}}
\begin{document}\cmsNoteHeader{EXO-23-015}

\newcommand{\tauprime}{{\HepParticle{\PGt}{}{\prime}}\xspace}
\newcommand{\tauprimep}{{\HepParticle{\tauprime}{}{+}}\xspace}
\newcommand{\tauprimem}{{\HepParticle{\tauprime}{}{-}}\xspace}
\newcommand{\atau}{\ensuremath{\mathrm{a}_{\PGt} }\xspace}
\newcommand{\actau}{\ensuremath{ c\tau_\mathrm{a}}\xspace}
\newcommand{\matau}{\ensuremath{ m_\mathrm{a} }\xspace}
\newcommand{\mtau}{\ensuremath{ m_{\PGt} }\xspace}
\newcommand{\mvll}{\ensuremath{ m_\mathrm{VLL} }\xspace}
\newcommand{\nhits}{\ensuremath{ N_\text{hits} }\xspace}
\newcommand{\tclusterCSC}{\ensuremath{ t_\text{cluster}^\mathrm{CSC} }\xspace}
\newcommand{\tspreadclusterCSC}{\ensuremath{ t_\text{spread}^\mathrm{CSC} }\xspace}
\newcommand{\tclusterDT}{\ensuremath{ t_\text{cluster}^\mathrm{DT} }\xspace}

\cmsNoteHeader{EXO-23-015}

\title{Search for vector-like leptons with long-lived particle decays in the CMS muon system in proton-proton collisions at \texorpdfstring{$\sqrt{s} = 13\TeV$}{sqrt(s) = 13 TeV}}

\date{\today}

\abstract{
    A first search is presented for vector-like leptons (VLLs) exclusively decaying into a light long-lived pseudoscalar boson and a standard model \PGt lepton. The pseudoscalar boson is assumed to have a mass below the \hbox{\PGtp\!\PGtm} threshold, so that it decays exclusively into two photons. It is identified using the CMS muon system. The analysis is carried out using a data set of proton-proton collisions at a center-of-mass energy of 13\TeV collected by the CMS experiment in 2016--2018, corresponding to an integrated luminosity of 138\fbinv. Selected events contain at least one pseudoscalar boson decaying electromagnetically in the muon system and at least one hadronically decaying \PGt lepton. No significant excess of data events is observed compared to the background expectation. Upper limits are set at 95\% confidence level on the vector-like lepton production cross section as a function of the VLL mass and the pseudoscalar boson mean proper decay length. The observed and expected exclusion ranges of the VLL mass extend up to 700 and 670\GeV, respectively, depending on the pseudoscalar boson lifetime.
}

\hypersetup{
    pdfauthor={CMS Collaboration},
    pdftitle={Search for vector-like leptons with long-lived particle decays in the CMS muon system in proton-proton collisions at sqrt(s) = 13 TeV},
    pdfsubject={CMS},
    pdfkeywords={vector-like leptons, tau, pseudoscalar, LLP, shower}}

\maketitle

\section{Introduction}
\label{sec:intro}

The standard model (SM) of particle physics successfully explains the interactions of elementary particles. However, some fundamental questions remain unexplained, such as the hierarchy problem and the nature of dark matter. One important feature of the SM is that fermions are chiral i.e., left- and right-handed fermions experience different weak interactions. Beyond-the-SM (BSM) models hypothesize the existence of additional chiral fermions, but direct searches for such fermions and measurements of the production cross section and properties of the Higgs boson (\PH) at the CERN LHC have largely excluded their existence~\cite{Djouadi:2012ae,Lenz:2013iha,ATLAS:2015lpr,CMS:2012wdt}. Therefore, if new elementary fermions exist, they should be of vector-like (nonchiral) nature. Vector-like fermions are among the simplest and well-motivated extensions of the SM~\cite{Aguilar-Saavedra:2017giu,Kumar:2015tna,Bhattiprolu:2019vdu}. They are predicted in various BSM scenarios, including, but not limited to, physics with extra dimensions~\cite{Kong:2010qd,Huang:2012kz}, string theories~\cite{Dijkstra:2004cc,Raby:2007hm,Lebedev:2006kn}, supersymmetric models~\cite{Martin:2009bg,Graham:2009gy,Endo:2011mc,Zheng:2019kqu}, and grand unification theories~\cite{Nevzorov:2012hs,Dorsner:2014wva,Joglekar:2016yap}.

At the LHC, vector-like quarks (VLQs) would be produced in pairs via quantum chromodynamics (QCD) processes or singly via electroweak interactions~\cite{Aguilar-Saavedra:2009xmz,Deandrea:2021vje}. Searches for VLQs have probed a large phase space, excluding mass hypotheses up to around 1.5\TeV~\cite{ATLAS:2018ziw,CMS:2022fck,CMS:2024bni}. In contrast, vector-like leptons (VLLs) would be produced via $s$-channel electroweak processes with much smaller production cross sections than the ones of VLQs. The VLL production cross section depends on the VLL mass, weak isospin, and weak hypercharge. The VLLs typically decay via mixing with the SM \PGt lepton into $\PGn\PW$, $\PGt\PZ$, or $\PGt\PH$~\cite{Kumar:2015tna,Bhattiprolu:2019vdu}. Alternative models predict more exotic decays into two SM third-generation quarks and one \PGt lepton via an off-shell leptoquark~\cite{Greljo:2018tuh}. Searches for VLLs have been recently performed by the ATLAS and CMS Collaborations using the data set of proton-proton ($\Pp\Pp$) collisions at $\sqrt{s}=13\TeV$, corresponding to an integrated luminosity of about 140\fbinv~\cite{ATLAS:2023sbu,ATLAS:2025wgc,CMS:2022nty,CMS:2022cpe,CMS:2024bni}. In particular, singlet VLLs with prompt decays in the mass range between 125 and 150\GeV are excluded by the CMS search~\cite{CMS:2022nty}.

In many BSM theories, vector-like fermions are accompanied by light pseudoscalar bosons, also referred to as axion-like particles (ALPs). Examples include composite Higgs models~\cite{Dobrescu:1999gv}, Little Higgs models~\cite{Cheng:2005as,Cacciapaglia:2009cv}, and theories of quark and lepton compositeness~\cite{Dobrescu:2021fny,Assi:2022jwg}. The presence of pseudoscalar bosons changes the collider phenomenology of the VLQs and VLLs and typically leads to final states that include low-mass diphoton resonances~\cite{Bauer:2017ris}.

\begin{figure*}[ht!]
    \centering
    \includegraphics[width=0.6\textwidth]{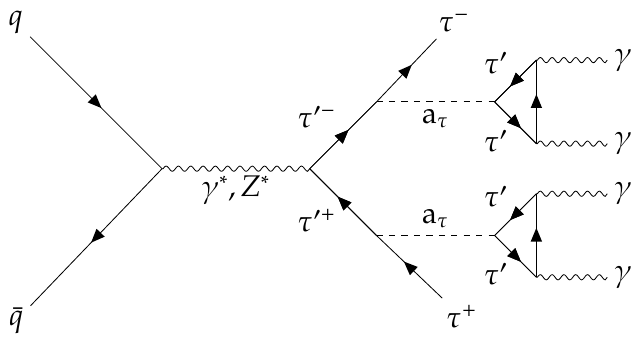}
    \caption{Feynman diagram of pair production of singlet vector-like leptons (\tauprime), which in turn both decay into an SM \PGt lepton and a new long-lived pseudoscalar boson ($\atau$).}
    \label{fig:vll_longlived_diagrams}
\end{figure*}

In this paper, a novel BSM physics model is studied in which a complex scalar field transforms as a singlet under the SU(2) gauge group and has Yukawa interactions with the weak-singlet VLL~\cite{Bernreuther:2023uxh}. The mass mixing between the VLL and the SM third-generation charged lepton leads to the emergence of a heavy physical VLL (\tauprime) and the well-established \PGt lepton. This complex scalar field has a vacuum expectation value and its charge-parity ($CP$)-odd component is lighter than $\tauprime$. The LHC phenomenology of this model is dominated by the VLL pair production $\Pp\Pp \to \tauprimep\tauprimem$ with subsequent $\tauprime$ decays to $\PGt\atau$, $\PGn \PW$, $\PGt \PZ$, and $\PGt \PH$, where $\atau$ is a new physical pseudoscalar boson. If the mass of the $\tauprime$ VLL ($\mvll$) arises from a Lagrangian term independent of the vacuum expectation value, then $\tauprime\rightarrow\PGt\atau$ is the dominant $\tauprime$ decay channel. For pseudoscalar boson masses ($\matau$) smaller than twice the \PGt mass ($\mtau$), $\atau$ predominantly decays to two photons (\PGg) via a $\tauprime$ loop. An example Feynman diagram of this process is presented in Fig.~\ref{fig:vll_longlived_diagrams}. Moreover, for typical values of parameters in this model (coupling between $\atau$ and $\tauprime$ of order 0.1, $\matau$ at the \GeV scale and $\mvll$ of order several hundreds of \GeV), $\atau$ will have a mean proper decay length ($\actau$) of a few tens of \unit{mm}~\cite{Bernreuther:2023uxh}. Given that $\atau$ is produced via the decay of a heavy particle, it typically has a very large momentum, thus $\atau$ may decay several meters away from the $\Pp\Pp$ interaction point.

This article presents the first search for singlet VLLs decaying into a light long-lived $\atau$ and a \PGt lepton. The subsequent decay of $\atau$ is assumed to be exclusively into two photons. The analysis is performed using a data set of $\Pp\Pp$ collisions at $\sqrt{s}=13\TeV$ collected by the CMS experiment from 2016 to 2018, collectively known as Run 2, and corresponding to an integrated luminosity of 138\fbinv. Events of interest contain at least one electromagnetically decaying $\atau$ in the muon system and at least one hadronically decaying \PGt lepton (\tauh). The collimated photons from the $\atau$ decay can produce a particle shower by interacting with the steel shielding material of the CMS muon system, similar to a sampling calorimeter. The shower is reconstructed as a localized high-multiplicity cluster of muon detector hits, denominated as a muon detector shower (MDS). The MDS object has been used in previous long-lived particle (LLP) searches by the CMS experiment~\cite{CMS:2021juv,CMS:2024bvl,CMS:2024ake}. Because of the calorimetric nature of the MDS signature, the MDS object is sensitive to the LLP energy rather than its mass. Therefore, this search would be equally sensitive to signal hypotheses for any $\matau<2\mtau$. A value of $\matau=2\GeV$ is used as a benchmark in this search. For $\matau>2\mtau$, $\atau$ would predominantly decay to \PGt leptons at tree level resulting in substantially shorter lifetimes~\cite{Bernreuther:2023uxh}. Thus alternative search strategies focusing on \PGt lepton final states could provide greater sensitivity for such scenarios. The VLL signal presented here (Fig. ~\ref{fig:vll_longlived_diagrams}) depends only on $\mvll$ and $\actau$. The results are interpreted within the VLL singlet model, although the search is also sensitive to VLL doublet models. Tabulated results and signal efficiencies are provided in the HEPData record for this analysis~\cite{hepdata}.

This paper is outlined as follows: the CMS detector is briefly described in Section~\ref{sec:cms}, followed by a description of the simulation samples in Section~\ref{sec:simulation}. Event reconstruction and selection are detailed in Sections~\ref{sec:reconstruction} and~\ref{sec:selection}, respectively. The background estimation strategy is described in Section~\ref{sec:background}. Sources of systematic uncertainties affecting the modeling of the signal and background are detailed in Section~\ref{sec:uncertainties}. The results are presented in Section~\ref{sec:results}, followed by a summary of this search in Section~\ref{sec:summary}.

\section{The CMS detector}
\label{sec:cms}

The central feature of the CMS detector is a superconducting solenoid of 6\unit{m} internal diameter, providing a magnetic field of 3.8\unit{T}. Within the solenoid volume are a silicon pixel and strip tracker, a lead tungstate crystal electromagnetic calorimeter (ECAL), and a brass and scintillator hadron calorimeter (HCAL), each composed of a barrel and two endcap sections. Forward calorimeters extend the pseudorapidity ($\eta$) coverage provided by the barrel and endcap detectors. Gas-ionization muon detectors embedded in the steel flux-return yoke outside the solenoid cover the region of $\abs{\eta}<2.4$. In Run 2, this system comprises three technologies: drift tubes (DTs), cathode strip chambers (CSCs), and resistive plate chambers (RPCs). More detailed descriptions of the CMS detector, together with a definition of the coordinate system used and the relevant kinematic variables, can be found in Refs.~\cite{CMS:2008xjf,CMS:2023gfb}.

The DT system covers the barrel region of $\abs{\eta}<1.2$. The DT chambers are arranged in five wheels along the beamline axis ($z$), and interleaved with layers of the steel flux-return yoke. Within each wheel, four concentric layers (stations) around the z axis are located approximately 4, 5, 6, and 7\unit{m} radially away from the interaction point, and named MB1--MB4 starting from the innermost one, where MB stands for ``muon barrel''. Each chamber consists of rectangular drift cells arranged in parallel and grouped in four layers to form superlayers (SLs). When charged particles transverse a cell, they ionize the gas and produce charges that drift towards an anode wire. The signal pulse measured is recorded as a hit. While all chambers are equipped with two SLs to measure the hit coordinate in the $r-\phi$ plane (where $\phi$ is azimuthal angle in radians), only MB1--MB3 station chambers contain a central SL to determine the hit z position.

The CSC system extends the coverage in the endcap region of $0.9<\abs{\eta}<2.4$. The CSC chambers are interleaved with the steel layers in four disks (stations) per endcap, located approximately 7.0, 8.0, 9.5, and 10.5\unit{m} away from the interaction point along the z axis. Each station consists of rings of chambers designated by MES/R, where ``ME'' means ``muon endcap'', S indicates the station number, and R is the ring number. The CSC chambers comprise six layers of cathode strips in the $r$ direction and anode wires perpendicular to the strip. Charged particles crossing a chamber ionize the gas and produce signal pulses in both anode wires and cathode strips. The information from anode and cathode signals are combined to provide the space and time coordinates of the hit in each layer.

The RPC detectors are double-gap chambers operated in avalanche mode. Signals are collected by readout strips, which are located in between the two gas gaps. The RPCs are characterized by their time resolution of around 1\unit{ns}, providing hits with a precise bunch-crossing assignment. They are mounted in both CSC and DT stations, covering the region of $\abs{\eta}<1.9$.

Events of interest are selected using a two-tiered trigger system. The first level, composed of custom hardware processors, uses information from the calorimeters and muon detectors to select events at a rate of around 100\unit{kHz} within a fixed latency of about 4\mus~\cite{CMS:2020cmk}. The second level, known as the high-level trigger, consists of a farm of processors running a version of the full event reconstruction software optimized for fast processing, and reduces the event rate to around 1\unit{kHz} before data storage~\cite{CMS:2016ngn,CMS:2024aqx}.

\section{Simulated samples}
\label{sec:simulation}

This search targets signal events with $\Pp\Pp\to\tauprimep\tauprimem$ production. The branching fractions $\mathcal{B}(\tauprime\to \PGt\atau)$ and $\mathcal{B}(\atau\to \PGg\PGg)$ are assumed to be 100\%. The hard scattering process is generated at leading order (LO) using the \MGvATNLO2.6.5 matrix element generator~\cite{MadGraph} and the VLL model presented in Ref.~\cite{Bernreuther:2023uxh}. Signal samples are generated with $\tauprime$ masses in the range of 200--800\GeV, where the trigger selection efficiencies are expected to be high, in steps of 100\GeV. A value of 2\GeV is taken for $\matau$. The parameter $\actau$ is set to various values in \MGvATNLO, ranging from 2\unit{mm} to 10\unit{m}. Since only a discrete set of $\actau$ values were generated, signal hypotheses with intermediate $\actau$ values are derived by reweighting the proper decay length distribution. Signal events are normalized to the next-to-LO (NLO) cross sections~\cite{Bernreuther:2023uxh} and the total integrated luminosity. Simulation is not used to estimate background events. The background estimate is derived using a purely data-driven method described in Section~\ref{sec:background}.

Generated events are interfaced with \PYTHIA8.230~\cite{Sj_strand_2015} to simulate parton showering and hadronization processes. Additional $\Pp\Pp$ interactions in the same or neighboring bunch crossings (pileup) are modeled by superimposing simulated minimum bias interactions onto the hard scattering process. Simulated events are reweighted to match the distribution of the number of pileup interactions observed during each data-taking period. The underlying event tune CP5~\cite{CP5} and NNPDF3.1~\cite{Ball:2017nwa} parton distribution functions are used in all simulated samples. All generated events are passed through a simulation of the CMS detector response based on the \GEANTfour toolkit~\cite{AGOSTINELLI2003250}.

\section{Event reconstruction}
\label{sec:reconstruction}

Reconstruction and identification of each particle in the event are performed by the particle-flow (PF) algorithm~\cite{CMS:2017yfk}, with an optimized combination of information from the various elements of the CMS detector. The energy of prompt photons that do not originate from the $\atau$ decay is obtained from the ECAL measurement. The energy of displaced di-photons from the $\atau$ decay is not measured by the MDS object. The energy of electrons is determined by a combination of the electron momentum at the primary interaction vertex as determined by the tracker, the energy of the corresponding ECAL cluster, and the energy sum of all bremsstrahlung photons spatially compatible with those originating from the electron track. The energy of muons is obtained from the curvature of the corresponding track. The energy of charged hadrons is determined from a combination of their momentum measured in the tracker and the matching ECAL and HCAL energy deposits, corrected for the response function of the calorimeters to hadronic showers. Finally, the energy of neutral hadrons is obtained from the corresponding corrected ECAL and HCAL energies.

The primary interaction vertex (PV) is defined as the vertex corresponding to the hardest scattering in the event, evaluated using only tracking information, as described in Section 9.4.1 of Ref.~\cite{CMS-TDR-15-02}.

Muons are measured in the region $\abs{\eta} < 2.4$, with detection planes comprising DTs, CSCs, and RPCs. The efficiency to reconstruct and identify muons is greater than 96\%. Matching muons to tracks measured in the silicon tracker results in a relative transverse momentum (\pt) resolutions of 1\% in the barrel and 3\% in the endcaps, for muons with $\pt < 100\GeV$; and better than 7\% in the barrel for muons with $\pt < 1\TeV$~\cite{CMS:2018rym}.

Hadronic jets are reconstructed by clustering PF candidates using the anti-\kt algorithm~\cite{Cacciari:2008gp, Cacciari:2011ma} with a distance parameter of 0.4. Pileup interactions can contribute with additional tracks and calorimetric energy depositions to the jet momentum. To mitigate this effect, charged particles identified to be originating from pileup vertices are discarded and an offset correction is applied to correct for remaining contributions~\cite{CMS:2020ebo}. Jet energy corrections are derived from simulation to bring the measured response of jets to that of particle-level jets on average. In situ measurements of the momentum balance in dijet, photon+jet, $\PZ$+jet, and multijet events are used to account for any residual differences in the jet energy scale between data and simulation~\cite{CMS:2016lmd}. The jet energy resolution typically is between 15 and 20\% at 30\GeV, 10\% at 100\GeV, and 5\% at 1\TeV~\cite{CMS:2016lmd}. Additional selection criteria are applied to each jet to remove jets potentially dominated by anomalous contributions from various subdetector components or reconstruction failures~\cite{CMS:2017wyc}.

For each event, the missing transverse momentum vector \ptvecmiss is computed as the negative vector \pt sum of all the PF candidates, and its magnitude is labeled as \ptmiss~\cite{CMS:2019ctu}. The \ptvecmiss is modified to account for corrections to the energy scale of the reconstructed jets in the event.

The \tauh candidates are reconstructed from jets, using the hadrons-plus-strips algorithm~\cite{CMS:2018jrd}, which combines one or three tracks with energy deposits in the calorimeters, to identify the \tauh decay modes.  Neutral pions are reconstructed as strips with dynamic size in $\eta$-$\phi$ from reconstructed electrons and photons, where the strip size varies as a function of the $\pt$ of the electron or photon candidate. Decay modes with one or three charged hadrons, with or without neutral pions, are considered in this analysis. The \textsc{DeepTau} algorithm is used~\cite{CMS:2022prd} to distinguish genuine \tauh decays from jets originating from the hadronization of quarks or gluons, and from electrons or muons. It combines the information from all individual reconstructed particles near the \tauh axis with properties of the \tauh candidate and the event. The jet misidentification rate of this algorithm depends on the \pt and quark flavor of the jet. In simulated events from \PW boson production in association with jets, it has been estimated to be 0.43\% for a genuine \tauh identification efficiency of 70\% at the chosen working point.

When LLPs decay within or just in front of the muon system, the material of the steel return yoke will induce a particle shower. Such showers will generate a localized and isolated cluster of muon detector hits. The CSC hits are reconstructed by combining the signal pulses in wire groups and cathode strips. The DT hits are reconstructed from signal pulses by the anode wires at the center of DT cells. Since DT chambers provide either the $\phi$ or $z$ coordinate, the DT hit position is assumed to be at the center of the DT chamber in the orthogonal direction.

\begin{figure*}[!htb]
    \centering
    {\includegraphics[width=1\columnwidth] {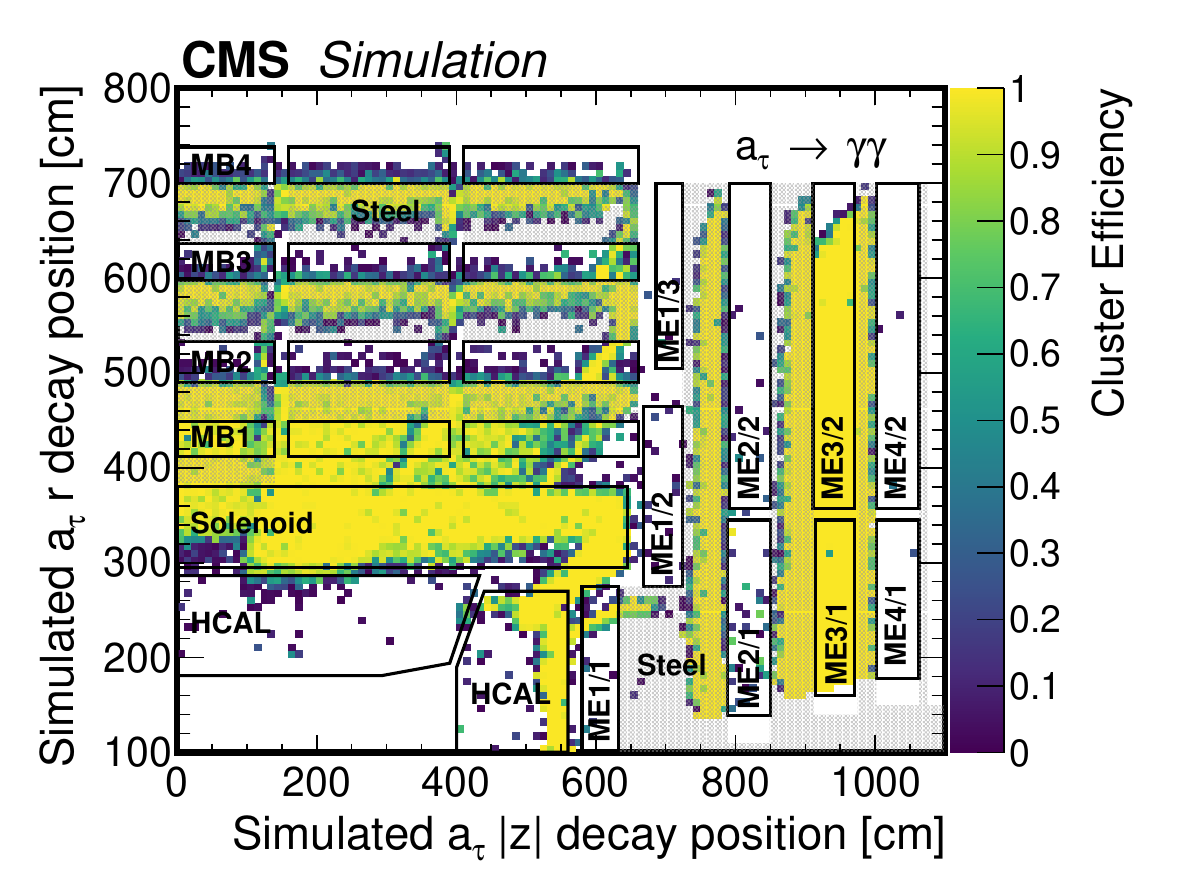}}
    \caption{\label{fig:clusterefficiency_mvll_700}
        The cluster reconstruction efficiency, including both DT and CSC clusters, as a function of the simulated $r$ and $\abs{z}$ decay positions of the pseudoscalar boson $\atau$ decaying to $\PGg\PGg$ in events with $\ptmiss>200\GeV$, for a VLL mass of 700\GeV, a pseudoscalar boson mass of 2\GeV, and a range of pseudoscalar boson mean proper decay lengths \actau uniformly distributed between 0.01 and 0.1\unit{m}. The barrel and endcap muon stations are drawn as black boxes and labeled by their names. The regions between muon stations are mostly composed of steel from the return yoke. The cluster reconstruction efficiency presented here corresponds to electromagnetic decays, in contrast to that of hadronic decays shown in Ref.~\cite{CMS:2024bvl}.}
\end{figure*}

The MDS objects are formed by clustering either CSC or DT hits based on their $\eta$ and $\phi$ coordinates using the density-based spatial clustering of applications with noise (DBSCAN algorithm~\cite{ref:dbscan}) with a distance parameter
$\Delta R = \sqrt{\smash[b]{(\Delta\eta)^2 + (\Delta\phi)^2}} = 0.2$. Given that minimum ionizing muons can produce up to 24 hits in the CSC and 44 hits in the DT system, a minimum of 50 hits per cluster are required in the DBSCAN algorithm. The efficiency to reconstruct an MDS object depends on the LLP decay mode (hadronic or electromagnetic). This analysis studies LLPs with electromagnetic decays. Consequently, particle showers are produced and contained at most within one station and stopped by the steel material between stations. When $\atau$ decays near or within the CSC or DT system, the inclusive cluster efficiency is approximately 35 and 45\%, respectively. The simulated cluster reconstruction efficiency, including both DT and CSC clusters, as a function of the $r$ and $\abs{z}$ decay positions of $\atau$ is shown in Fig.~\ref{fig:clusterefficiency_mvll_700}. The modeling of the MDS object in simulation has been studied using a sample enriched with $\PZ\to\PGmp\PGmm$, where one of the muons emits a photon through bremsstrahlung. The photon produces an electromagnetic shower in the muon system, and thus it is reconstructed as an MDS cluster. Differences in data and simulation efficiencies are accounted for as systematic uncertainties described in Section~\ref{sec:uncertainties}.

\section{Event selection}
\label{sec:selection}

This analysis focuses on signal events with candidates for at least one \tauh and at least one $\atau$ decay in the muon system. Given that the $\atau$ decays beyond the calorimeters, its momentum is not properly accounted for by the PF algorithm and a large value of $\ptmiss$ is reconstructed in the event. We exploit this feature and analyze the data collected by triggers with online $\ptmiss>120\GeV$. After trigger selection, we impose a minimum offline $\ptmiss$ of 200\GeV to be above the threshold where the $\ptmiss$ trigger becomes fully efficient. The \ptmiss signal selection efficiency, calculated assumming at least one $\atau$ decays within the geometric acceptance, ranges from 6 to 62\%, depending on the $\mvll$ mass hypothesis.

We require events to have at least one reconstructed \tauh candidate with $\pt > 30\GeV$, $\abs{\eta}<2.3$, satisfying the identification criteria described in Section~\ref{sec:reconstruction}. The \tauh selection efficiency is between 50 and 85\% depending on the signal hypothesis, with around 98\% background rejection.

Selected events are required to contain at least one MDS cluster. They are categorized into DT- and CSC-cluster categories, based on the cluster type. There is only one cluster selected per event. If a DT cluster and a CSC cluster are present in the event, the event will be assigned to the DT category. Both categories have similar SM background sources that could induce a particle shower in the muon system, and therefore produce an MDS cluster. Ordered by relative importance, the background sources are the following: (i) energetic particles from quark/gluon hadronization or secondary interactions not stopped by the shielding material, known as jet punch-through; (ii) muons that undergo bremsstrahlung in the muon system; (iii) isolated hadrons from pileup or underlying events (single particle punch-through); and (iv) cosmic ray muon showers. To reject these backgrounds, additional requirements are applied to MDS clusters, as detailed in the following subsections.

\subsection{The DT-cluster category}
\label{sec:selectiondt}
The DT-cluster category targets signal events in which the $\atau$ is reconstructed as a DT cluster. To suppress punch-through jet and muon bremsstrahlung backgrounds, we veto clusters with a centroid within $\Delta R=0.4$ of a jet with $\pt> 15\GeV$ and $\abs{\eta}<3.0$ or a muon with $\pt> 15\GeV$ and $\abs{\eta}<2.4$. In addition, we observe shower events induced by cosmic rays where a large number of hits are observed in multiple muon chambers in adjacent wheels. These are suppressed by vetoing clusters containing more than 8 MB1 hits within $\Delta\phi<\pi/4$ in any of the wheels adjacent to the wheel containing the cluster hits.

We also require the cluster to coincide with at least one RPC detector hit in the same wheel and within $\Delta\phi<0.5$. The time measurements of those RPC hits are matched to discrete bunch crossing times. Then, the DT cluster time ($\tclusterDT$) is defined as the most frequently occurring bunch crossing time among the matched RPC hits. Lastly, the DT cluster time is required to match the PV bunch crossing time ($\tclusterDT=0$), to reject clusters produced from pileup or interactions from nearby bunch crossings, denominated out-of-time (OOT) pileup.

\subsection{The CSC-cluster category}
The CSC-cluster category targets signal events in which the $\atau$ is reconstructed as a CSC cluster. The jet and muon vetoes described in Section~\ref{sec:selectiondt} are applied to the CSC cluster. Clusters matched to a DT segment or RPC hit in the MB1 station within $\Delta R<0.4$ are also rejected in order to remove clusters produced from muons traversing the DT region $0.8 <\abs{\eta}<1.2$ and showering in the CSC chambers. Clusters with an ME1/1 hit are removed because of the large punch-through jet background due to the reduced shielding present in this region. Finally, clusters with $\abs{\eta}>2.2$ are rejected to suppress muon bremsstrahlung background not accounted for by the muon veto due to inefficient muon reconstruction.

The CSC cluster time ($\tclusterCSC$) and time spread ($t_\text{spread}^\mathrm{CSC}$) are defined as the average and root-mean-square spread of the timestamps of the hits comprising the cluster, respectively. To reject clusters from the OOT pileup background, $-5.0<\tclusterCSC<12.5$\unit{ns} and $\tspreadclusterCSC<20$\unit{ns} are required.

\section{Background modeling}
\label{sec:background}

Signal LLPs are expected to produce large numbers of hits in the muon system compared to background processes. Therefore, we use the distribution of number of hits in the cluster ($\nhits$) as the signal discriminant in both DT- and CSC-cluster categories. The signal region (SR) is defined as events passing the selections detailed in Section~\ref{sec:selection}. The SR background estimate is derived from data events in a signal-depleted region.

\begin{figure}[!htb]
    \centering
    \includegraphics[width=0.497\columnwidth]{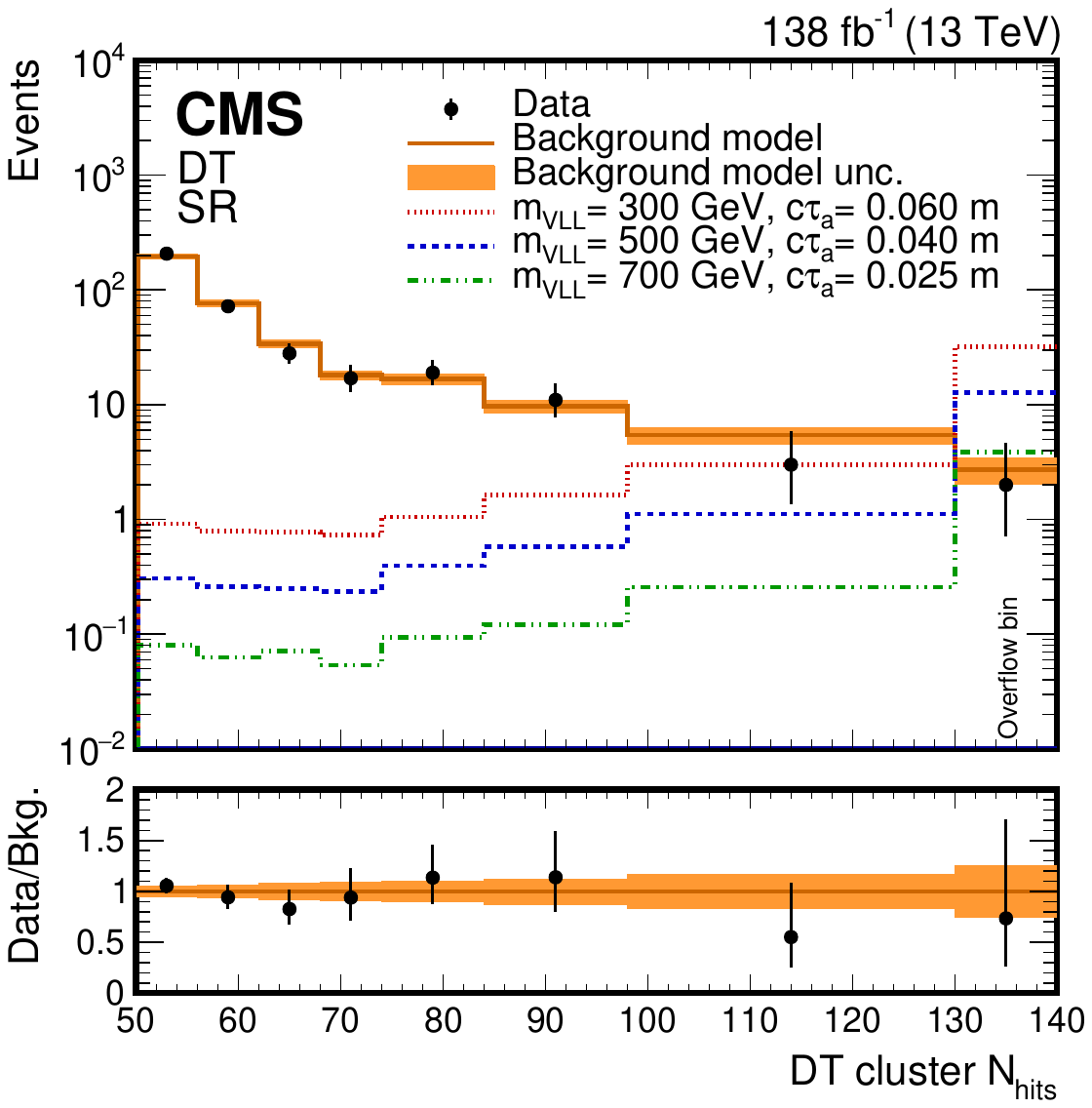}
    \includegraphics[width=0.497\columnwidth]{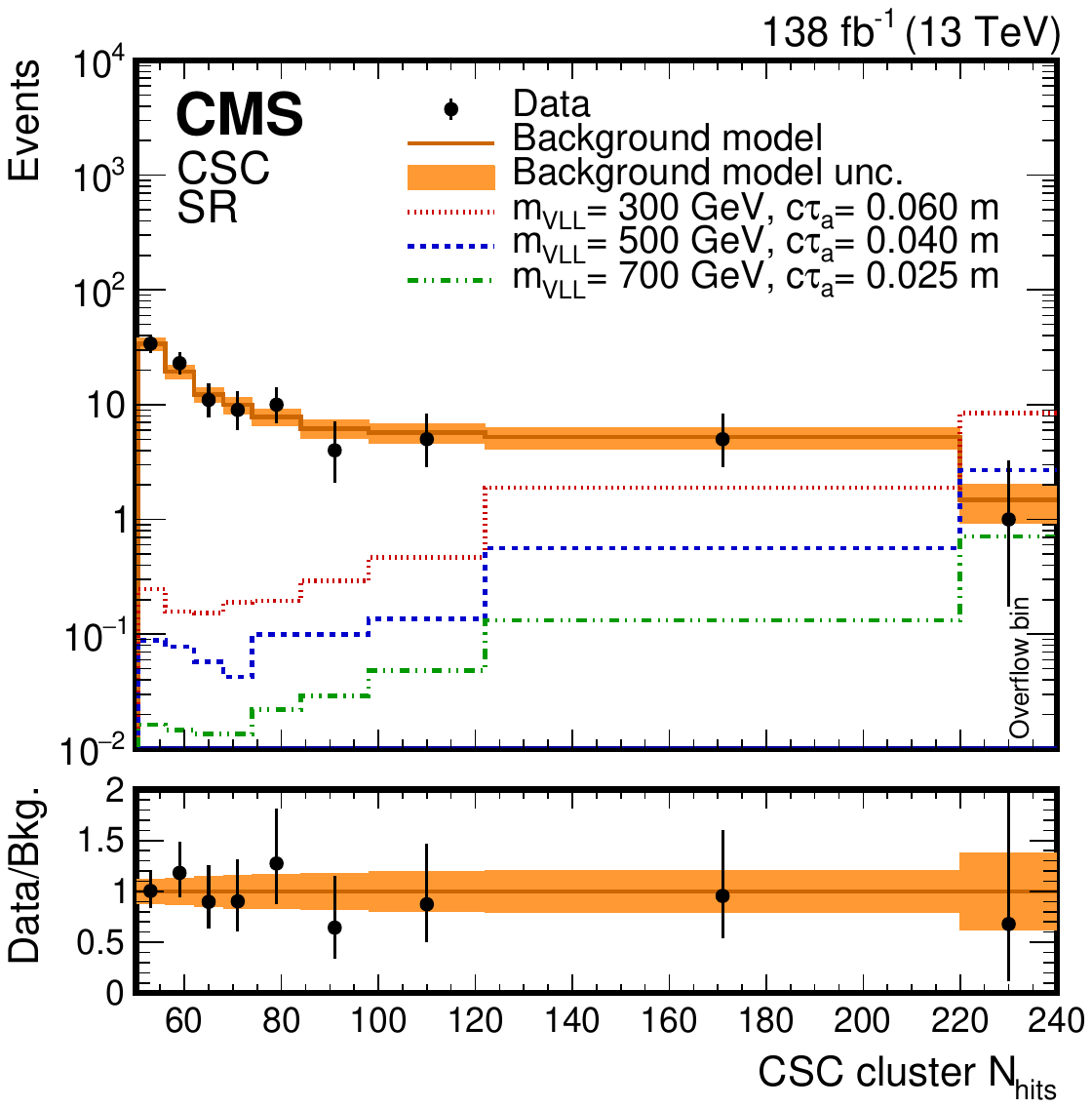} \\
    \caption{Distributions of the number of hits in the cluster ($\nhits$) for the DT (left) and  CSC (right) cluster categories in the signal region (SR). The black markers represent the data. The solid orange line and associated orange band show the background prediction and corresponding uncertainty. The red dotted, blue dashed and green dashed-dotted lines denote different signal hypotheses for a pseudoscalar boson with a mass of 2\GeV. The last histogram bin contains all overflow events. The lower panel in each plot shows the ratio of the data to the estimated background.}
    \label{fig:sr_rechits_cscdt_intime}
\end{figure}

Since the cluster properties are expected to be independent of the \tauh selection, a control region (CR) is defined by requiring at least one reconstructed \tauh candidate but inverting the \tauh \textsc{DeepTau} identification requirements with respect to the ones applied to the SR events. The expected background $\nhits$ distribution shape is taken from the CR and normalized to the data measured in the SR with a constant scaling factor. Fisher F-tests~\cite{ref:ftest} and saturated model goodness-of-fit tests~\cite{Baker:1983tu,lindsey1996parametric} have been performed establishing that a higher-order polynomial dependence, beyond a constant scaling factor between the SR and CR distributions as a function of $\nhits$, is not needed. The SR $\nhits$ distributions are presented in Fig.~\ref{fig:sr_rechits_cscdt_intime}.

\begin{figure}[!htb]
    \centering
    \includegraphics[width=0.497\columnwidth]{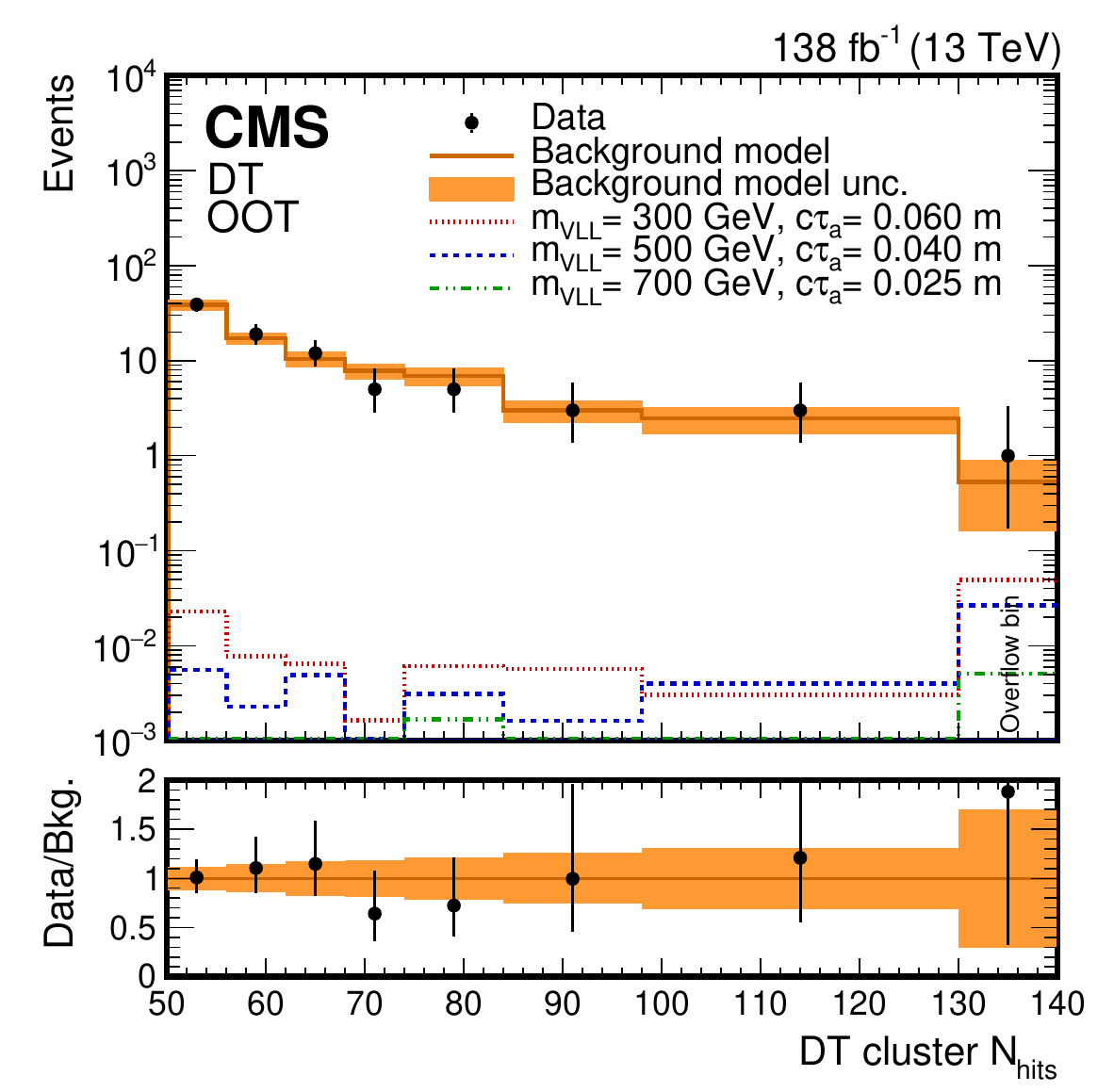}
    \includegraphics[width=0.497\columnwidth]{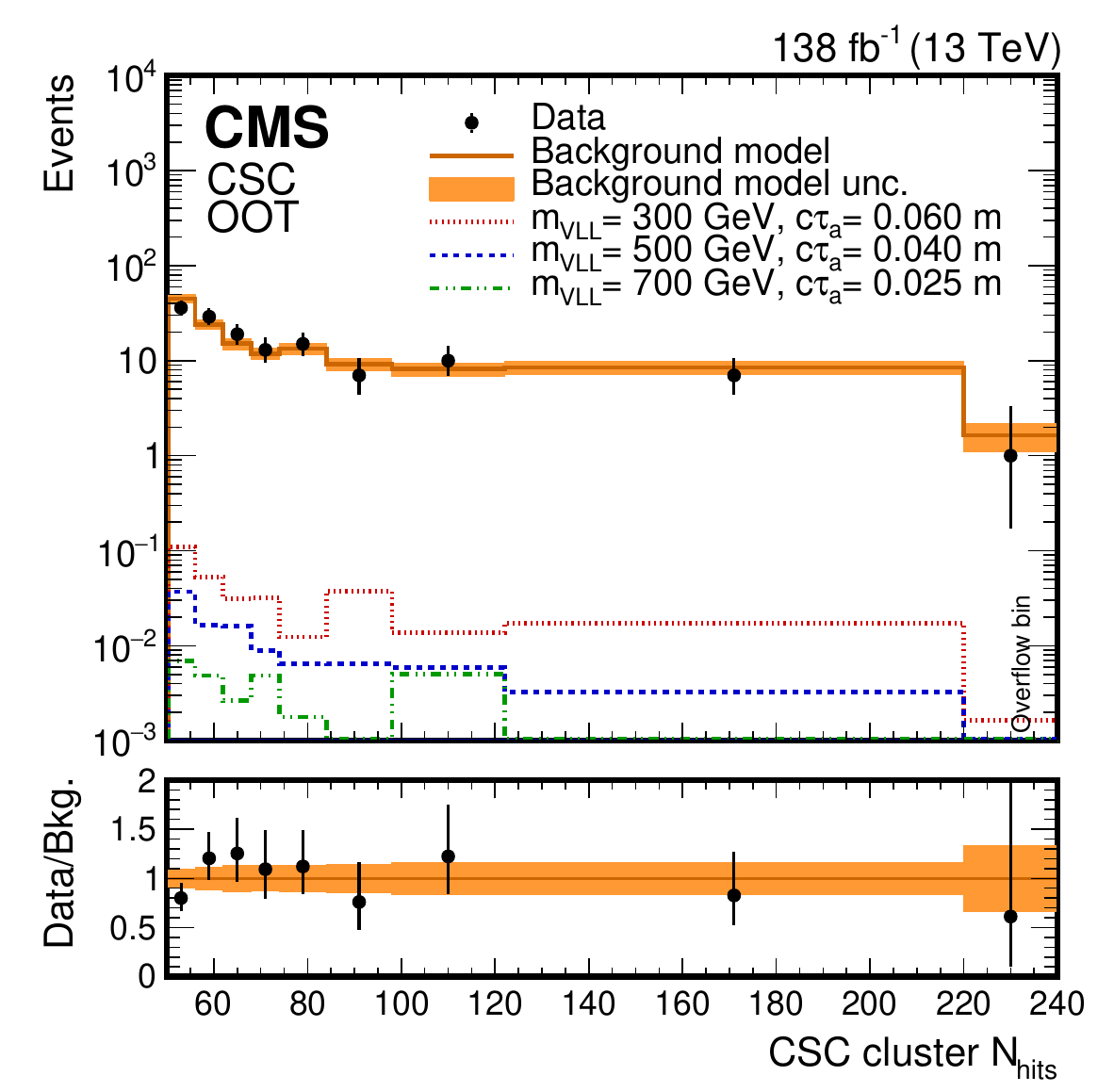}
    \caption{Distributions of the number of hits in the cluster ($\nhits$) for the DT (left) and  CSC (right) cluster categories in the out-of-time (OOT) region. The black markers represent the data. The solid orange line and associated orange band show the background prediction and corresponding uncertainty. The red dotted, blue dashed and green dashed-dotted lines denote different signal hypotheses for a pseudoscalar boson with a mass of 2\GeV. The last histogram bin contains all overflow events. The lower panel in each plot shows the ratio of the data to the estimated background.}
    \label{fig:sr_rechits_cscdt_ootime}
\end{figure}

This background model is validated in OOT validation regions that are orthogonal to the SR. The DT and CSC OOT regions contain events passing all selection criteria, except that a negative cluster time is required: $\tclusterDT<0$ and $\tclusterCSC<-12.5$\unit{ns}, respectively. In analogy to the SR, the shape of the expected background $\nhits$ distribution is taken from the CR and normalized to data measured in the OOT with a constant scaling factor. Statistical tests are performed and show that a constant scaling factor is sufficient to model the background distribution in both DT and CSC OOT validation regions. The validation OOT distributions of $\nhits$ are shown in Fig.~\ref{fig:sr_rechits_cscdt_ootime}.

\section{Systematic uncertainties}
\label{sec:uncertainties}

The background uncertainty arises from the statistical fluctuation in the CR data. Since good agreement is observed between the background estimation and data in the OOT validation regions, no additional uncertainty is considered. Background uncertainties are uncorrelated across categories.

The simulation of the muon detector response in an environment with a large multiplicity of secondary particles is validated by studying data and simulation in a region with $\PZ\to\PGmp\PGmm$ events, where a muon undergoes bremsstrahlung and the associated photon produces an MDS cluster. Uncertainties in the CSC and DT cluster reconstruction efficiency are estimated to be around 13 and 16\%, respectively. Simulation correction factors are also derived to account for the mismodeling of cluster selection efficiencies, and their uncertainties are propagated to the final result. For CSC clusters, a 6.8\% correction factor is applied to the muon veto efficiency, with a systematic uncertainty of 4.5\%. The uncertainties in the CSC cluster time spread and cluster time selections are 2.8 and 0.9\%, respectively. In addition, a 1\% uncertainty is assigned to account for differences in the CSC readout conditions in simulation and data, which could lead to an overestimation of the $\nhits$ in the cluster.

The \tauh identification efficiencies in simulation are corrected to match those in data. The uncertainties of these correction factors are propagated to the final result as uncertainties in the expected signal yield (5--11\%). Lastly, we also account for experimental uncertainties in the signal yield prediction: pileup (1\%), \tauh energy scale (1\%), jet energy scale (1--4\%), trigger efficiency (1--4\%), and total integrated luminosity (1.6\%)~\cite{CMS-LUM-17-003,CMS-PAS-LUM-17-004,CMS-PAS-LUM-18-002}. Theoretical uncertainties in the signal modeling are found to have a negligible impact on the analysis. The signal systematic uncertainties, except for the ones in the cluster modeling, are assumed to be correlated across categories.

\section{Results}
\label{sec:results}
No statistically significant deviation from a background-only hypothesis is observed. We evaluate the upper limits on the VLL production cross section using the modified frequentist \CLs~\cite{JUNK1999435,Read_2002} criterion with the profile likelihood ratio as the test statistic~\cite{CMS-NOTE-2011-005}. The \CLs value is calculated using an asymptotic approximation of the test statistic~\cite{Cowan:2010js}. The sources of systematic uncertainties detailed in Section~\ref{sec:uncertainties} are modeled by log-normal distributions in the likelihood function. The final results are obtained by a simultaneous fit of the $\nhits$ distributions in both DT- and CSC-cluster categories using the CMS statistical analysis tool \textsc{Combine}~\cite{CMS:2024onh}, which is based on the \textsc{RooFit}~\cite{Verkerke:2003ir} and \textsc{RooStats}~\cite{Moneta:2010pm} frameworks. The DT-cluster category contributes the most to the overall sensitivity due to its higher signal-to-background ratio. The exclusion limits obtained using the asymptotic approximation are consistent within 10\% with those based on pseudo-experiments.

\begin{figure}[!htb]
    \centering
    \includegraphics[width=0.497\columnwidth]{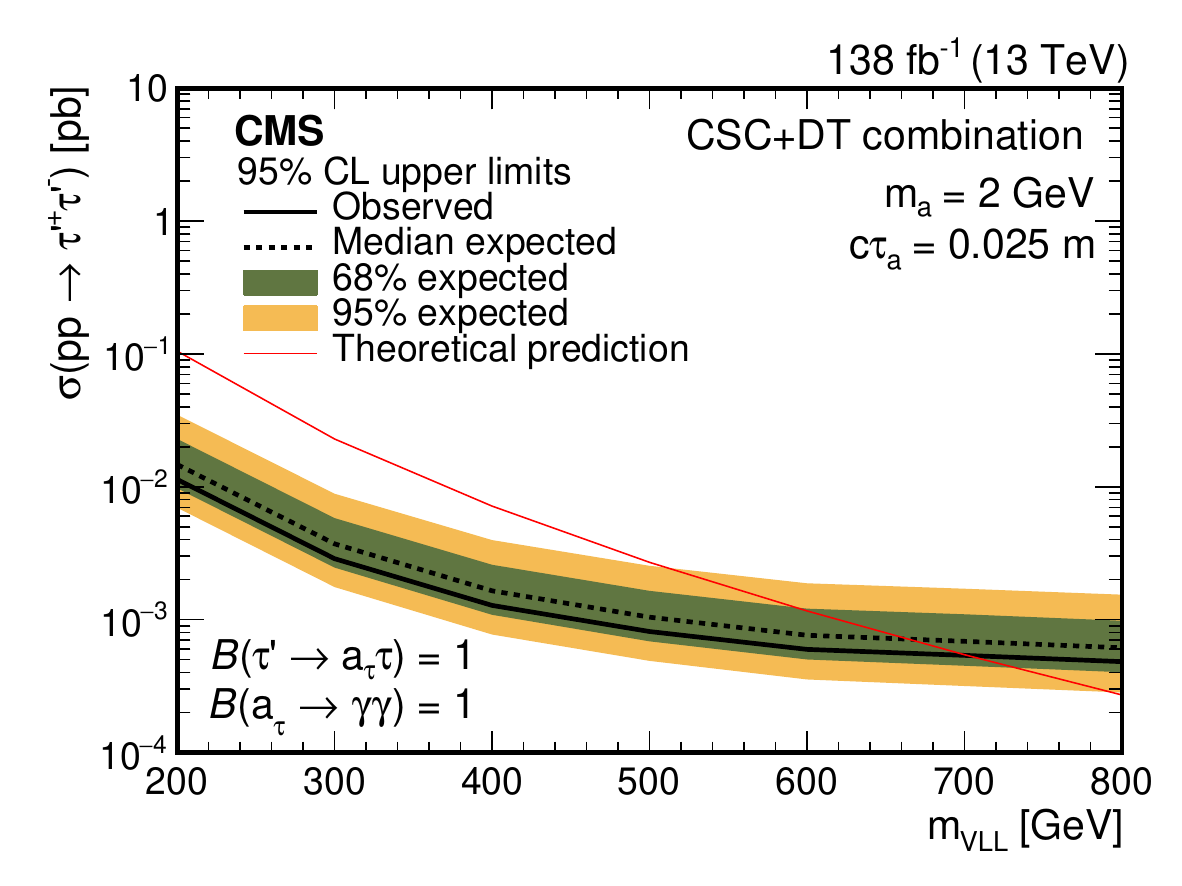}
    \includegraphics[width=0.497\columnwidth]{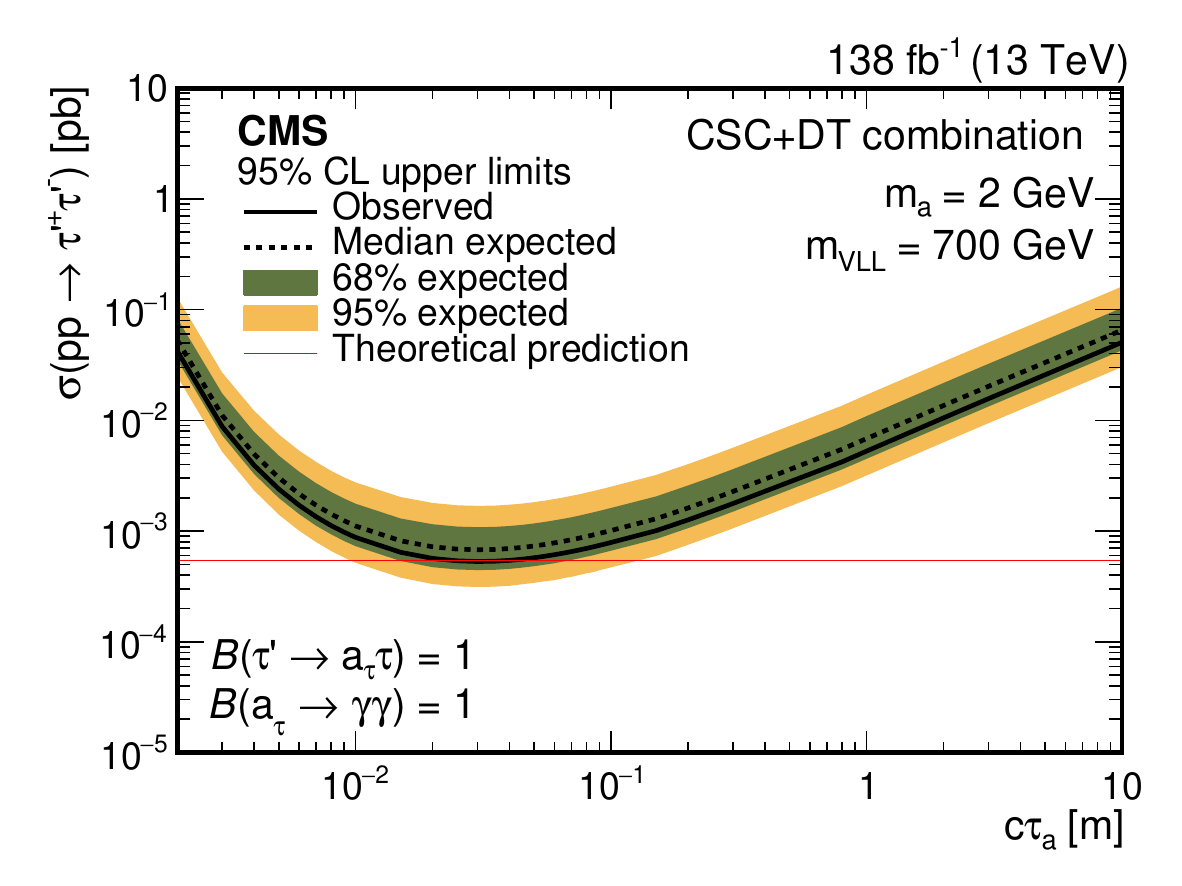}
    \caption{The 95\% \CL observed and expected upper limits on the VLL production cross section as a function of the VLL mass for the pseudoscalar boson mean proper decay length $\actau=0.025$\unit{m} (left), and as a function of $\actau$ for VLL mass of 700\GeV (right). The pseudoscalar boson mass is 2\GeV. The NLO theoretical prediction~\cite{Bernreuther:2023uxh} is shown as a red line.}
    \label{fig:results_1d_limits}
\end{figure}

\begin{figure}[!htb]
    \centering
    \includegraphics[width=0.75\columnwidth]{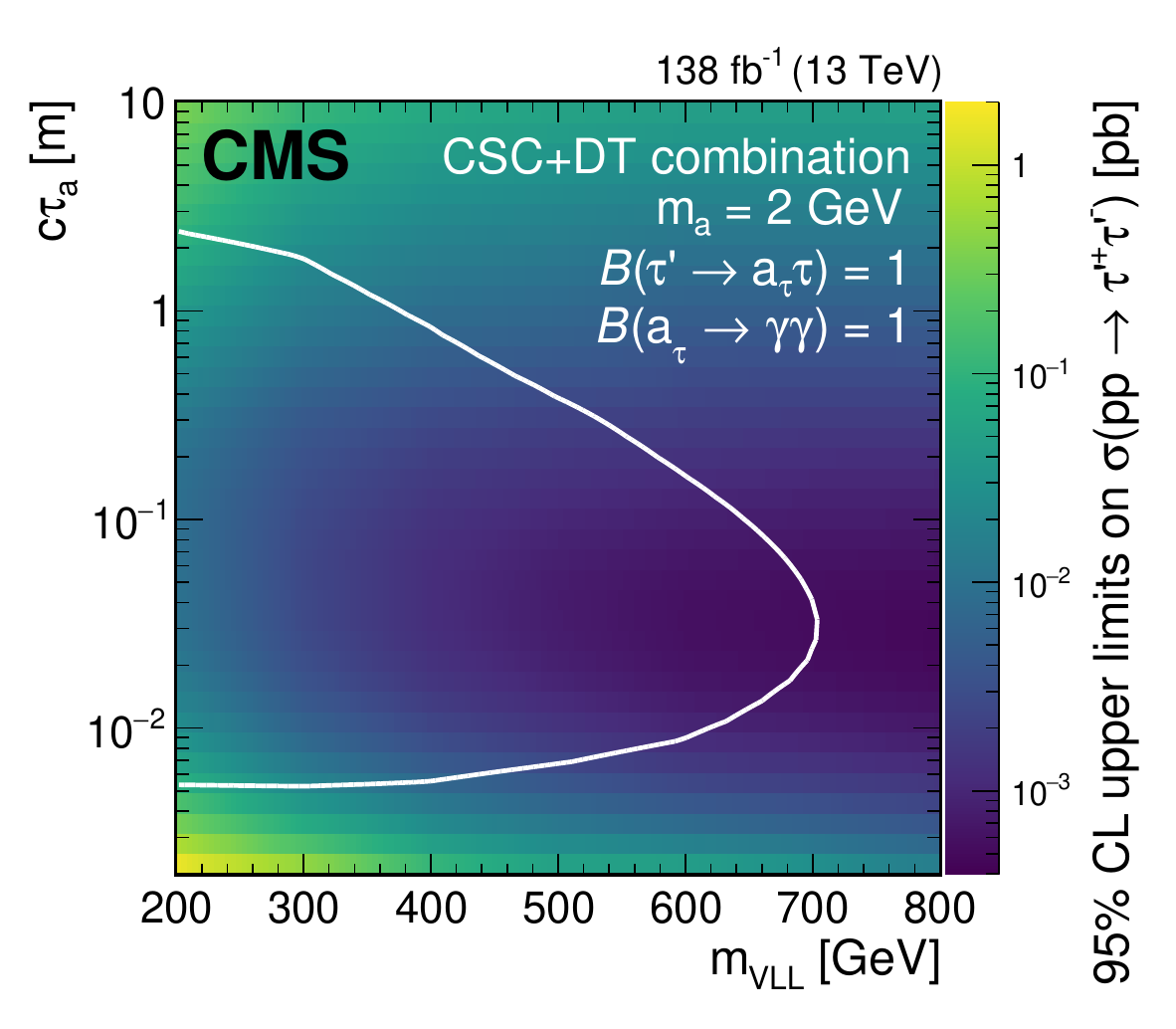}
    \caption{The 95\% \CL observed upper limits on the VLL production cross section as a function of the VLL mass and the pseudoscalar boson mean proper decay length $\actau$. The pseudoscalar boson mass is 2\GeV. The area enclosed by the white line corresponds to the excluded region.}
    \label{fig:results_2d_limits}
\end{figure}

The observed and expected upper limits at 95\% confidence level (\CL) on the VLL cross section are presented as a function of VLL mass and $\actau$ in Fig.~\ref{fig:results_1d_limits}. In this search, VLL masses below 700\GeV are observed to be excluded with 670\GeV expected, depending on $\actau$. Figure~\ref{fig:results_2d_limits} shows the observed limit as a function of both the VLL mass and $\actau$.


\section{Summary}
\label{sec:summary}

The first search for singlet vector-like leptons (VLLs) that decay into a light long-lived pseudoscalar boson and a \PGt lepton has been presented. It is performed using the CMS data set of proton-proton collisions at 13\TeV collected in 2016--2018, corresponding to an integrated luminosity of 138\fbinv. This analysis targets a reconstructed signature with at least one hadronically decaying tau lepton and with at least one muon detector shower resulting from the pseudoscalar boson decay in the CMS muon system. Selected events are categorized based on the presence of a cluster of muon detector hits in the barrel or the endcap region. No significant deviation from the background-only hypothesis is observed. The results of each category are combined to derive upper limits on the VLL production cross section as a function of the VLL mass and the pseudoscalar boson proper decay length. For pseudoscalar boson mean proper decay lengths in the range of 0.005--2.4 \unit{m}, VLL masses up to 700\GeV are excluded at 95\% confidence level. These are the most stringent constraints on the production of singlet VLLs with long-lived decays.

\begin{acknowledgments}
    We congratulate our colleagues in the CERN accelerator departments for the excellent performance of the LHC and thank the technical and administrative staffs at CERN and at other CMS institutes for their contributions to the success of the CMS effort. In addition, we gratefully acknowledge the computing centers and personnel of the Worldwide LHC Computing Grid and other centers for delivering so effectively the computing infrastructure essential to our analyses. Finally, we acknowledge the enduring support for the construction and operation of the LHC, the CMS detector, and the supporting computing infrastructure provided by the following funding agencies: SC (Armenia), BMBWF and FWF (Austria); FNRS and FWO (Belgium); CNPq, CAPES, FAPERJ, FAPERGS, and FAPESP (Brazil); MES and BNSF (Bulgaria); CERN; CAS, MoST, and NSFC (China); MINCIENCIAS (Colombia); MSES and CSF (Croatia); RIF (Cyprus); SENESCYT (Ecuador); ERC PRG, RVTT3 and MoER TK202 (Estonia); Academy of Finland, MEC, and HIP (Finland); CEA and CNRS/IN2P3 (France); SRNSF (Georgia); BMBF, DFG, and HGF (Germany); GSRI (Greece); NKFIH (Hungary); DAE and DST (India); IPM (Iran); SFI (Ireland); INFN (Italy); MSIP and NRF (Republic of Korea); MES (Latvia); LMTLT (Lithuania); MOE and UM (Malaysia); BUAP, CINVESTAV, CONACYT, LNS, SEP, and UASLP-FAI (Mexico); MOS (Montenegro); MBIE (New Zealand); PAEC (Pakistan); MES and NSC (Poland); FCT (Portugal); MESTD (Serbia); MICIU/AEI and PCTI (Spain); MOSTR (Sri Lanka); Swiss Funding Agencies (Switzerland); MST (Taipei); MHESI and NSTDA (Thailand); TUBITAK and TENMAK (Turkey); NASU (Ukraine); STFC (United Kingdom); DOE and NSF (USA).

    \hyphenation{Rachada-pisek} Individuals have received support from the Marie-Curie program and the European Research Council and Horizon 2020 Grant, contract Nos.\ 675440, 724704, 752730, 758316, 765710, 824093, 101115353, 101002207, and COST Action CA16108 (European Union); the Leventis Foundation; the Alfred P.\ Sloan Foundation; the Alexander von Humboldt Foundation; the Science Committee, project no. 22rl-037 (Armenia); the Fonds pour la Formation \`a la Recherche dans l'Industrie et dans l'Agriculture (FRIA-Belgium); the Beijing Municipal Science \& Technology Commission, No. Z191100007219010 and Fundamental Research Funds for the Central Universities (China); the Ministry of Education, Youth and Sports (MEYS) of the Czech Republic; the Shota Rustaveli National Science Foundation, grant FR-22-985 (Georgia); the Deutsche Forschungsgemeinschaft (DFG), among others, under Germany's Excellence Strategy -- EXC 2121 ``Quantum Universe" -- 390833306, and under project number 400140256 - GRK2497; the Hellenic Foundation for Research and Innovation (HFRI), Project Number 2288 (Greece); the Hungarian Academy of Sciences, the New National Excellence Program - \'UNKP, the NKFIH research grants K 131991, K 133046, K 138136, K 143460, K 143477, K 146913, K 146914, K 147048, 2020-2.2.1-ED-2021-00181, TKP2021-NKTA-64, and 2021-4.1.2-NEMZ\_KI-2024-00036 (Hungary); the Council of Science and Industrial Research, India; ICSC -- National Research Center for High Performance Computing, Big Data and Quantum Computing and FAIR -- Future Artificial Intelligence Research, funded by the NextGenerationEU program (Italy); the Latvian Council of Science; the Ministry of Education and Science, project no. 2022/WK/14, and the National Science Center, contracts Opus 2021/41/B/ST2/01369 and 2021/43/B/ST2/01552 (Poland); the Funda\c{c}\~ao para a Ci\^encia e a Tecnologia, grant CEECIND/01334/2018 (Portugal); the National Priorities Research Program by Qatar National Research Fund; MICIU/AEI/10.13039/501100011033, ERDF/EU, "European Union NextGenerationEU/PRTR", and Programa Severo Ochoa del Principado de Asturias (Spain); the Chulalongkorn Academic into Its 2nd Century Project Advancement Project, and the National Science, Research and Innovation Fund via the Program Management Unit for Human Resources \& Institutional Development, Research and Innovation, grant B39G670016 (Thailand); the Kavli Foundation; the Nvidia Corporation; the SuperMicro Corporation; the Welch Foundation, contract C-1845; and the Weston Havens Foundation (USA).
\end{acknowledgments}
\bibliography{auto_generated}

\providecommand{\href}[2]{#2}\begingroup\raggedright\begin{thebibliography}{10}%
\makeatletter
\providecommand{\hrefCMSnoop }[0]{\@secondoftwo}%
\makeatother
\providecommand{\doi}{\texttt{doi:}\begingroup \urlstyle{tt}\Url}

\bibitem{Djouadi:2012ae}
\hrefCMSnoop {}{A.~Djouadi and A.~Lenz, ``{Sealing the fate of a fourth
  generation of fermions}'',} \textit{ Phys. Lett. B} \textbf{ 715} (2012) 310,
  \href{http://dx.doi.org/10.1016/j.physletb.2012.07.060}{\doi{10.1016/j.physletb.2012.07.060}},
  \href{http://www.arXiv.org/abs/1204.1252}{\texttt{arXiv:1204.1252}}.

\bibitem{Lenz:2013iha}
\hrefCMSnoop {}{A.~Lenz, ``{Constraints on a fourth generation of fermions from
  Higgs boson searches}'',} \textit{ Adv. High Energy Phys.} \textbf{ 2013}
  (2013) 910275,
  \href{http://dx.doi.org/10.1155/2013/910275}{\doi{10.1155/2013/910275}}.

\bibitem{ATLAS:2015lpr}
\hrefCMSnoop {}{{ATLAS Collaboration}, ``{Search for pair production of a new
  heavy quark that decays into a $W$ boson and a light quark in $\Pp\Pp$
  collisions at $\sqrt{s} = 8\TeV$ with the ATLAS detector}'',} \textit{ Phys.
  Rev. D} \textbf{ 92} (2015) 112007,
  \href{http://dx.doi.org/10.1103/PhysRevD.92.112007}{\doi{10.1103/PhysRevD.92.112007}},
  \href{http://www.arXiv.org/abs/1509.04261}{\texttt{arXiv:1509.04261}}.

\bibitem{CMS:2012wdt}
\hrefCMSnoop {}{{CMS Collaboration}, ``{Combined search for the quarks of a
  sequential fourth generation}'',} \textit{ Phys. Rev. D} \textbf{ 86} (2012)
  112003,
  \href{http://dx.doi.org/10.1103/PhysRevD.86.112003}{\doi{10.1103/PhysRevD.86.112003}},
  \href{http://www.arXiv.org/abs/1209.1062}{\texttt{arXiv:1209.1062}}.

\bibitem{Aguilar-Saavedra:2017giu}
\hrefCMSnoop {}{J.~A. Aguilar-Saavedra, D.~E. L{\'o}pez-Fogliani, and
  C.~Mu{\~n}oz, ``{Novel signatures for vector-like quarks}'',} \textit{ JHEP}
  \textbf{ 06} (2017) 095,
  \href{http://dx.doi.org/10.1007/JHEP06(2017)095}{\doi{10.1007/JHEP06(2017)095}},
  \href{http://www.arXiv.org/abs/1705.02526}{\texttt{arXiv:1705.02526}}.

\bibitem{Kumar:2015tna}
\hrefCMSnoop {}{N.~Kumar and S.~P. Martin, ``{Vectorlike leptons at the Large
  Hadron Collider}'',} \textit{ Phys. Rev. D} \textbf{ 92} (2015) 115018,
  \href{http://dx.doi.org/10.1103/PhysRevD.92.115018}{\doi{10.1103/PhysRevD.92.115018}},
  \href{http://www.arXiv.org/abs/1510.03456}{\texttt{arXiv:1510.03456}}.

\bibitem{Bhattiprolu:2019vdu}
\hrefCMSnoop {}{P.~N. Bhattiprolu and S.~P. Martin, ``{Prospects for vectorlike
  leptons at future proton-proton colliders}'',} \textit{ Phys. Rev. D}
  \textbf{ 100} (2019) 015033,
  \href{http://dx.doi.org/10.1103/PhysRevD.100.015033}{\doi{10.1103/PhysRevD.100.015033}},
  \href{http://www.arXiv.org/abs/1905.00498}{\texttt{arXiv:1905.00498}}.

\bibitem{Kong:2010qd}
\hrefCMSnoop {}{K.~Kong, S.~C. Park, and T.~G. Rizzo, ``{A vector-like fourth
  generation with a discrete symmetry from Split-UED}'',} \textit{ JHEP}
  \textbf{ 07} (2010) 059,
  \href{http://dx.doi.org/10.1007/JHEP07(2010)059}{\doi{10.1007/JHEP07(2010)059}},
  \href{http://www.arXiv.org/abs/1004.4635}{\texttt{arXiv:1004.4635}}.

\bibitem{Huang:2012kz}
\hrefCMSnoop {}{G.-Y. Huang, K.~Kong, and S.~C. Park, ``{Bounds on the
  fermion-bulk masses in models with universal extra dimensions}'',} \textit{
  JHEP} \textbf{ 06} (2012) 099,
  \href{http://dx.doi.org/10.1007/JHEP06(2012)099}{\doi{10.1007/JHEP06(2012)099}},
  \href{http://www.arXiv.org/abs/1204.0522}{\texttt{arXiv:1204.0522}}.

\bibitem{Dijkstra:2004cc}
\hrefCMSnoop {}{T.~P.~T. Dijkstra, L.~R. Huiszoon, and A.~N. Schellekens,
  ``{Supersymmetric standard model spectra from RCFT orientifolds}'',} \textit{
  Nucl. Phys. B} \textbf{ 710} (2005) 3,
  \href{http://dx.doi.org/10.1016/j.nuclphysb.2004.12.032}{\doi{10.1016/j.nuclphysb.2004.12.032}},
  \href{http://www.arXiv.org/abs/hep-th/0411129}{\texttt{arXiv:hep-th/0411129}}.

\bibitem{Raby:2007hm}
\hrefCMSnoop {}{S.~Raby and A.~Wingerter, ``{Gauge coupling unification and
  light exotica in string theory}'',} \textit{ Phys. Rev. Lett.} \textbf{ 99}
  (2007) 051802,
  \href{http://dx.doi.org/10.1103/PhysRevLett.99.051802}{\doi{10.1103/PhysRevLett.99.051802}},
  \href{http://www.arXiv.org/abs/0705.0294}{\texttt{arXiv:0705.0294}}.

\bibitem{Lebedev:2006kn}
O.~Lebedev\hrefCMSnoop {}{ { et~al.}, ``{A mini-landscape of exact MSSM spectra
  in heterotic orbifolds}'',} \textit{ Phys. Lett. B} \textbf{ 645} (2007) 88,
  \href{http://dx.doi.org/10.1016/j.physletb.2006.12.012}{\doi{10.1016/j.physletb.2006.12.012}},
  \href{http://www.arXiv.org/abs/hep-th/0611095}{\texttt{arXiv:hep-th/0611095}}.

\bibitem{Martin:2009bg}
\hrefCMSnoop {}{S.~P. Martin, ``{Extra vector-like matter and the lightest
  Higgs scalar boson mass in low-energy supersymmetry}'',} \textit{ Phys. Rev.
  D} \textbf{ 81} (2010) 035004,
  \href{http://dx.doi.org/10.1103/PhysRevD.81.035004}{\doi{10.1103/PhysRevD.81.035004}},
  \href{http://www.arXiv.org/abs/0910.2732}{\texttt{arXiv:0910.2732}}.

\bibitem{Graham:2009gy}
\hrefCMSnoop {}{P.~W. Graham, A.~Ismail, S.~Rajendran, and P.~Saraswat, ``A
  little solution to the little hierarchy problem: {A} vector-like
  generation'',} \textit{ Phys. Rev. D} \textbf{ 81} (2010) 055016,
  \href{http://dx.doi.org/10.1103/PhysRevD.81.055016}{\doi{10.1103/PhysRevD.81.055016}},
  \href{http://www.arXiv.org/abs/0910.3020}{\texttt{arXiv:0910.3020}}.

\bibitem{Endo:2011mc}
\hrefCMSnoop {}{M.~Endo, K.~Hamaguchi, S.~Iwamoto, and N.~Yokozaki, ``{Higgs}
  mass and muon anomalous magnetic moment in supersymmetric models with
  vector-like matters'',} \textit{ Phys. Rev. D} \textbf{ 84} (2011) 075017,
  \href{http://dx.doi.org/10.1103/PhysRevD.84.075017}{\doi{10.1103/PhysRevD.84.075017}},
  \href{http://www.arXiv.org/abs/1108.3071}{\texttt{arXiv:1108.3071}}.

\bibitem{Zheng:2019kqu}
\hrefCMSnoop {}{S.~Zheng, ``Minimal vectorlike model in supersymmetric
  unification'',} \textit{ Eur. Phys. J. C} \textbf{ 80} (2020) 273,
  \href{http://dx.doi.org/10.1140/epjc/s10052-020-7843-8}{\doi{10.1140/epjc/s10052-020-7843-8}},
  \href{http://www.arXiv.org/abs/1904.10145}{\texttt{arXiv:1904.10145}}.

\bibitem{Nevzorov:2012hs}
\hrefCMSnoop {}{R.~Nevzorov, ``{$E_6$ inspired supersymmetric models with exact
  custodial symmetry}'',} \textit{ Phys. Rev. D} \textbf{ 87} (2013) 015029,
  \href{http://dx.doi.org/10.1103/PhysRevD.87.015029}{\doi{10.1103/PhysRevD.87.015029}},
  \href{http://www.arXiv.org/abs/1205.5967}{\texttt{arXiv:1205.5967}}.

\bibitem{Dorsner:2014wva}
\hrefCMSnoop {}{I.~Dor{\v{s}}ner, S.~Fajfer, and I.~Musta{\'c}, ``{Light
  vector-like fermions in a minimal SU(5) setup}'',} \textit{ Phys. Rev. D}
  \textbf{ 89} (2014) 115004,
  \href{http://dx.doi.org/10.1103/PhysRevD.89.115004}{\doi{10.1103/PhysRevD.89.115004}},
  \href{http://www.arXiv.org/abs/1401.6870}{\texttt{arXiv:1401.6870}}.

\bibitem{Joglekar:2016yap}
\hrefCMSnoop {}{A.~Joglekar and J.~L. Rosner, ``{Searching for signatures of
  $E_{6}$}'',} \textit{ Phys. Rev. D} \textbf{ 96} (2017) 015026,
  \href{http://dx.doi.org/10.1103/PhysRevD.96.015026}{\doi{10.1103/PhysRevD.96.015026}},
  \href{http://www.arXiv.org/abs/1607.06900}{\texttt{arXiv:1607.06900}}.

\bibitem{Aguilar-Saavedra:2009xmz}
\hrefCMSnoop {}{J.~A. Aguilar-Saavedra, ``{Identifying top partners at LHC}'',}
  \textit{ JHEP} \textbf{ 11} (2009) 030,
  \href{http://dx.doi.org/10.1088/1126-6708/2009/11/030}{\doi{10.1088/1126-6708/2009/11/030}},
  \href{http://www.arXiv.org/abs/0907.3155}{\texttt{arXiv:0907.3155}}.

\bibitem{Deandrea:2021vje}
A.~Deandrea\hrefCMSnoop {}{ { et~al.}, ``{Single production of vector-like
  quarks: The effects of large width, interference and NLO corrections}'',}
  \textit{ JHEP} \textbf{ 08} (2021) 107,
  \href{http://dx.doi.org/10.1007/JHEP08(2021)107}{\doi{10.1007/JHEP08(2021)107}},
  \href{http://www.arXiv.org/abs/2105.08745}{\texttt{arXiv:2105.08745}}.
  [Erratum: DOI{10.1007/JHEP11(2022)028}].

\bibitem{ATLAS:2018ziw}
\hrefCMSnoop {}{{ATLAS Collaboration}, ``{Combination of the searches for
  pair-produced vector-like partners of the third-generation quarks at
  $\sqrt{s} =13\TeV$ with the ATLAS detector}'',} \textit{ Phys. Rev. Lett.}
  \textbf{ 121} (2018) 211801,
  \href{http://dx.doi.org/10.1103/PhysRevLett.121.211801}{\doi{10.1103/PhysRevLett.121.211801}},
  \href{http://www.arXiv.org/abs/1808.02343}{\texttt{arXiv:1808.02343}}.

\bibitem{CMS:2022fck}
\hrefCMSnoop {}{{CMS Collaboration}, ``{Search for pair production of
  vector-like quarks in leptonic final states in proton-proton collisions at
  $\sqrt{s} = 13\TeV$}'',} \textit{ JHEP} \textbf{ 07} (2023) 020,
  \href{http://dx.doi.org/10.1007/JHEP07(2023)020}{\doi{10.1007/JHEP07(2023)020}},
  \href{http://www.arXiv.org/abs/2209.07327}{\texttt{arXiv:2209.07327}}.

\bibitem{CMS:2024bni}
\hrefCMSnoop {}{{CMS Collaboration}, ``{Review of searches for vector-like
  quarks, vector-like leptons, and heavy neutral leptons in proton-proton
  collisions at $\sqrt{s} = 13\TeV$ at the CMS experiment}'',} \textit{ Phys.
  Rept.} \textbf{ 1115} (2025) 570,
  \href{http://dx.doi.org/10.1016/j.physrep.2024.09.012}{\doi{10.1016/j.physrep.2024.09.012}},
  \href{http://www.arXiv.org/abs/2405.17605}{\texttt{arXiv:2405.17605}}.

\bibitem{Greljo:2018tuh}
\hrefCMSnoop {}{A.~Greljo and B.~A. Stefanek, ``{Third family quark-lepton
  unification at the \TeVns scale}'',} \textit{ Phys. Lett. B} \textbf{ 782}
  (2018) 131,
  \href{http://dx.doi.org/10.1016/j.physletb.2018.05.033}{\doi{10.1016/j.physletb.2018.05.033}},
  \href{http://www.arXiv.org/abs/1802.04274}{\texttt{arXiv:1802.04274}}.

\bibitem{ATLAS:2023sbu}
\hrefCMSnoop {}{{ATLAS Collaboration}, ``{Search for third-generation
  vector-like leptons in $\Pp\Pp$ collisions at $\sqrt{s} = 13\TeV$ with the
  ATLAS detector}'',} \textit{ JHEP} \textbf{ 07} (2023) 118,
  \href{http://dx.doi.org/10.1007/JHEP07(2023)118}{\doi{10.1007/JHEP07(2023)118}},
  \href{http://www.arXiv.org/abs/2303.05441}{\texttt{arXiv:2303.05441}}.

\bibitem{ATLAS:2025wgc}
\hrefCMSnoop {}{{ATLAS Collaboration}, ``{Search for electroweak production of
  vector-like leptons in $\tau$-lepton and $\PQb$-jet final states in $\Pp\Pp$
  collisions at $\sqrt{s} = 13\TeV$ with the ATLAS detector}'',} 2025.
  \href{http://www.arXiv.org/abs/2503.22581}{\texttt{arXiv:2503.22581}}.
  Submitted to \emph{Eur. Phys. J. C}.

\bibitem{CMS:2022nty}
\hrefCMSnoop {}{{CMS Collaboration}, ``{Inclusive nonresonant multilepton
  probes of new phenomena at $\sqrt{s}=13\TeV$}'',} \textit{ Phys. Rev. D}
  \textbf{ 105} (2022) 112007,
  \href{http://dx.doi.org/10.1103/PhysRevD.105.112007}{\doi{10.1103/PhysRevD.105.112007}},
  \href{http://www.arXiv.org/abs/2202.08676}{\texttt{arXiv:2202.08676}}.

\bibitem{CMS:2022cpe}
\hrefCMSnoop {}{{CMS Collaboration}, ``{Search for pair-produced vector-like
  leptons in final states with third-generation leptons and at least three \PQb
  quark jets in proton-proton collisions at $\sqrt{s} = 13\TeV$}'',} \textit{
  Phys. Lett. B} \textbf{ 846} (2023) 137713,
  \href{http://dx.doi.org/10.1016/j.physletb.2023.137713}{\doi{10.1016/j.physletb.2023.137713}},
  \href{http://www.arXiv.org/abs/2208.09700}{\texttt{arXiv:2208.09700}}.

\bibitem{Dobrescu:1999gv}
\hrefCMSnoop {}{B.~A. Dobrescu, ``{Minimal composite Higgs model with light
  bosons}'',} \textit{ Phys. Rev. D} \textbf{ 63} (2001) 015004,
  \href{http://dx.doi.org/10.1103/PhysRevD.63.015004}{\doi{10.1103/PhysRevD.63.015004}},
  \href{http://www.arXiv.org/abs/hep-ph/9908391}{\texttt{arXiv:hep-ph/9908391}}.

\bibitem{Cheng:2005as}
\hrefCMSnoop {}{H.-C. Cheng, I.~Low, and L.-T. Wang, ``{Top partners in little
  Higgs theories with T-parity}'',} \textit{ Phys. Rev. D} \textbf{ 74} (2006)
  055001,
  \href{http://dx.doi.org/10.1103/PhysRevD.74.055001}{\doi{10.1103/PhysRevD.74.055001}},
  \href{http://www.arXiv.org/abs/hep-ph/0510225}{\texttt{arXiv:hep-ph/0510225}}.

\bibitem{Cacciapaglia:2009cv}
\hrefCMSnoop {}{G.~Cacciapaglia, A.~Deandrea, S.~R. Choudhury, and N.~Gaur,
  ``{T-parity odd heavy leptons at LHC}'',} \textit{ Phys. Rev. D} \textbf{ 81}
  (2010) 075005,
  \href{http://dx.doi.org/10.1103/PhysRevD.81.075005}{\doi{10.1103/PhysRevD.81.075005}},
  \href{http://www.arXiv.org/abs/0911.4632}{\texttt{arXiv:0911.4632}}.

\bibitem{Dobrescu:2021fny}
\hrefCMSnoop {}{B.~A. Dobrescu, ``Quark and lepton compositeness: A
  renormalizable model'',} \textit{ Phys. Rev. Lett.} \textbf{ 128} (2022)
  241804,
  \href{http://dx.doi.org/10.1103/PhysRevLett.128.241804}{\doi{10.1103/PhysRevLett.128.241804}},
  \href{http://www.arXiv.org/abs/2112.15132}{\texttt{arXiv:2112.15132}}.

\bibitem{Assi:2022jwg}
\hrefCMSnoop {}{B.~Assi and B.~A. Dobrescu, ``{Proton decay from quark and
  lepton compositeness}'',} \textit{ JHEP} \textbf{ 12} (2022) 116,
  \href{http://dx.doi.org/10.1007/JHEP12(2022)116}{\doi{10.1007/JHEP12(2022)116}},
  \href{http://www.arXiv.org/abs/2211.02211}{\texttt{arXiv:2211.02211}}.

\bibitem{Bauer:2017ris}
\hrefCMSnoop {}{M.~Bauer, M.~Neubert, and A.~Thamm, ``Collider probes of
  axion-like particles'',} \textit{ JHEP} \textbf{ 12} (2017) 044,
  \href{http://dx.doi.org/10.1007/JHEP12(2017)044}{\doi{10.1007/JHEP12(2017)044}},
  \href{http://www.arXiv.org/abs/1708.00443}{\texttt{arXiv:1708.00443}}.

\bibitem{Bernreuther:2023uxh}
\hrefCMSnoop {}{E.~Bernreuther and B.~A. Dobrescu, ``{Vectorlike leptons and
  long-lived bosons at the LHC}'',} \textit{ JHEP} \textbf{ 07} (2023) 079,
  \href{http://dx.doi.org/10.1007/JHEP07(2023)079}{\doi{10.1007/JHEP07(2023)079}},
  \href{http://www.arXiv.org/abs/2304.08509}{\texttt{arXiv:2304.08509}}.

\bibitem{CMS:2021juv}
\hrefCMSnoop {}{{CMS Collaboration}, ``{Search for long-lived particles
  decaying in the CMS endcap muon detectors in proton-proton collisions at
  $\sqrt{s} =13\TeV$}'',} \textit{ Phys. Rev. Lett.} \textbf{ 127} (2021)
  261804,
  \href{http://dx.doi.org/10.1103/PhysRevLett.127.261804}{\doi{10.1103/PhysRevLett.127.261804}},
  \href{http://www.arXiv.org/abs/2107.04838}{\texttt{arXiv:2107.04838}}.

\bibitem{CMS:2024bvl}
\hrefCMSnoop {}{{CMS Collaboration}, ``{Search for long-lived particles
  decaying in the CMS muon detectors in proton-proton collisions at
  $\sqrt{s}=13\TeV$}'',} \textit{ Phys. Rev. D} \textbf{ 110} (2024) 032007,
  \href{http://dx.doi.org/10.1103/PhysRevD.110.032007}{\doi{10.1103/PhysRevD.110.032007}},
  \href{http://www.arXiv.org/abs/2402.01898}{\texttt{arXiv:2402.01898}}.

\bibitem{CMS:2024ake}
\hrefCMSnoop {}{{CMS Collaboration}, ``{Search for long-lived heavy neutral
  leptons decaying in the CMS muon detectors in proton-proton collisions at
  $\sqrt{s}=13\TeV$}'',} \textit{ Phys. Rev. D} \textbf{ 110} (2024) 012004,
  \href{http://dx.doi.org/10.1103/PhysRevD.110.012004}{\doi{10.1103/PhysRevD.110.012004}},
  \href{http://www.arXiv.org/abs/2402.18658}{\texttt{arXiv:2402.18658}}.

\bibitem{hepdata}
\hrefCMSnoop {}{}{HEPD}ata record for this analysis, 2025.
\newblock
  \href{http://dx.doi.org/10.17182/hepdata.156846}{\doi{10.17182/hepdata.156846}}.

\bibitem{CMS:2008xjf}
\hrefCMSnoop {}{{CMS Collaboration}, ``The {CMS} experiment at the {CERN}
  {LHC}'',} \textit{ JINST} \textbf{ 3} (2008) S08004,
  \href{http://dx.doi.org/10.1088/1748-0221/3/08/S08004}{\doi{10.1088/1748-0221/3/08/S08004}}.

\bibitem{CMS:2023gfb}
\hrefCMSnoop {}{{CMS Collaboration}, ``{Development of the CMS detector for the
  CERN LHC Run 3}'',} \textit{ JINST} \textbf{ 19} (2024) P05064,
  \href{http://dx.doi.org/10.1088/1748-0221/19/05/P05064}{\doi{10.1088/1748-0221/19/05/P05064}},
  \href{http://www.arXiv.org/abs/2309.05466}{\texttt{arXiv:2309.05466}}.

\bibitem{CMS:2020cmk}
\hrefCMSnoop {}{{CMS Collaboration}, ``{Performance of the CMS level-1 trigger
  in proton-proton collisions at $\sqrt{s} = 13\TeV$}'',} \textit{ JINST}
  \textbf{ 15} (2020) P10017,
  \href{http://dx.doi.org/10.1088/1748-0221/15/10/P10017}{\doi{10.1088/1748-0221/15/10/P10017}},
  \href{http://www.arXiv.org/abs/2006.10165}{\texttt{arXiv:2006.10165}}.

\bibitem{CMS:2016ngn}
\hrefCMSnoop {}{{CMS Collaboration}, ``{The CMS trigger system}'',} \textit{
  JINST} \textbf{ 12} (2017) P01020,
  \href{http://dx.doi.org/10.1088/1748-0221/12/01/P01020}{\doi{10.1088/1748-0221/12/01/P01020}},
\href{http://www.arXiv.org/abs/1609.02366}{\texttt{arXiv:1609.02366}}.

\bibitem{CMS:2024aqx}
\hrefCMSnoop {}{{CMS Collaboration}}
  \href{http://dx.doi.org/10.1088/1748-0221/19/11/P11021}{\doi{10.1088/1748-0221/19/11/P11021}}.

\bibitem{MadGraph}
J.~Alwall\hrefCMSnoop {}{ { et~al.}, ``The automated computation of tree-level
  and next-to-leading order differential cross sections, and their matching to
  parton shower simulations'',} \textit{ JHEP} \textbf{ 07} (2014) 079,
  \href{http://dx.doi.org/10.1007/JHEP07(2014)079}{\doi{10.1007/JHEP07(2014)079}},
  \href{http://www.arXiv.org/abs/1405.0301}{\texttt{arXiv:1405.0301}}.

\bibitem{Sj_strand_2015}
T.~Sj{\"o}strand\hrefCMSnoop {}{ { et~al.}, ``An introduction to
  {\PYTHIA8.2}'',} \textit{ Comp. Phys. Comm.} \textbf{ 191} (2015) 159,
  \href{http://dx.doi.org/10.1016/j.cpc.2015.01.024}{\doi{10.1016/j.cpc.2015.01.024}},
  \href{http://www.arXiv.org/abs/1410.3012}{\texttt{arXiv:1410.3012}}.

\bibitem{CP5}
\hrefCMSnoop {}{{CMS Collaboration}, ``Extraction and validation of a new set
  of {CMS \PYTHIA}8 tunes from underlying-event measurements'',} \textit{ Eur.
  Phys. J. C} \textbf{ 80} (2020) 4,
  \href{http://dx.doi.org/10.1140/epjc/s10052-019-7499-4}{\doi{10.1140/epjc/s10052-019-7499-4}},
  \href{http://www.arXiv.org/abs/1903.12179}{\texttt{arXiv:1903.12179}}.

\bibitem{Ball:2017nwa}
\hrefCMSnoop {}{{NNPDF} Collaboration, ``Parton distributions from
  high-precision collider data'',} \textit{ Eur. Phys. J. C} \textbf{ 77}
  (2017) 663,
  \href{http://dx.doi.org/10.1140/epjc/s10052-017-5199-5}{\doi{10.1140/epjc/s10052-017-5199-5}},
  \href{http://www.arXiv.org/abs/1706.00428}{\texttt{arXiv:1706.00428}}.

\bibitem{AGOSTINELLI2003250}
\hrefCMSnoop {}{{GEANT4} Collaboration, ``{\GEANTfour}---a simulation
  toolkit'',} \textit{ Nucl. Instrum. Meth. A} \textbf{ 506} (2003) 250,
  \href{http://dx.doi.org/10.1016/S0168-9002(03)01368-8}{\doi{10.1016/S0168-9002(03)01368-8}}.

\bibitem{CMS:2017yfk}
\hrefCMSnoop {}{{CMS Collaboration}, ``Particle-flow reconstruction and global
  event description with the {CMS} detector'',} \textit{ JINST} \textbf{ 12}
  (2017) P10003,
  \href{http://dx.doi.org/10.1088/1748-0221/12/10/P10003}{\doi{10.1088/1748-0221/12/10/P10003}},
\href{http://www.arXiv.org/abs/1706.04965}{\texttt{arXiv:1706.04965}}.

\bibitem{CMS-TDR-15-02}
\href {http://cds.cern.ch/record/2020886}{{CMS Collaboration}, ``Technical
  proposal for the {Phase-II} upgrade of the {Compact Muon Solenoid}'',} CMS
  Technical Proposal CERN-LHCC-2015-010, CMS-TDR-15-02, 2015.

\bibitem{CMS:2018rym}
\hrefCMSnoop {}{{CMS Collaboration}, ``{Performance of the CMS muon detector
  and muon reconstruction with proton-proton collisions at $\sqrt{s}=
  13\TeV$}'',} \textit{ JINST} \textbf{ 13} (2018) P06015,
  \href{http://dx.doi.org/10.1088/1748-0221/13/06/P06015}{\doi{10.1088/1748-0221/13/06/P06015}},
  \href{http://www.arXiv.org/abs/1804.04528}{\texttt{arXiv:1804.04528}}.

\bibitem{Cacciari:2008gp}
\hrefCMSnoop {}{M.~Cacciari, G.~P. Salam, and G.~Soyez, ``{The anti-$\kt$ jet
  clustering algorithm}'',} \textit{ JHEP} \textbf{ 04} (2008) 063,
  \href{http://dx.doi.org/10.1088/1126-6708/2008/04/063}{\doi{10.1088/1126-6708/2008/04/063}},
  \href{http://www.arXiv.org/abs/0802.1189}{\texttt{arXiv:0802.1189}}.

\bibitem{Cacciari:2011ma}
\hrefCMSnoop {}{M.~Cacciari, G.~P. Salam, and G.~Soyez, ``{FastJet} user
  manual'',} \textit{ Eur. Phys. J. C} \textbf{ 72} (2012) 1896,
  \href{http://dx.doi.org/10.1140/epjc/s10052-012-1896-2}{\doi{10.1140/epjc/s10052-012-1896-2}},
  \href{http://www.arXiv.org/abs/1111.6097}{\texttt{arXiv:1111.6097}}.

\bibitem{CMS:2020ebo}
\hrefCMSnoop {}{{CMS Collaboration}, ``{Pileup mitigation at CMS in 13\TeV
  data}'',} \textit{ JINST} \textbf{ 15} (2020) P09018,
  \href{http://dx.doi.org/10.1088/1748-0221/15/09/P09018}{\doi{10.1088/1748-0221/15/09/P09018}},
  \href{http://www.arXiv.org/abs/2003.00503}{\texttt{arXiv:2003.00503}}.

\bibitem{CMS:2016lmd}
\hrefCMSnoop {}{{CMS Collaboration}, ``{Jet energy scale and resolution in the
  CMS experiment in $\Pp\Pp$ collisions at 8\TeV}'',} \textit{ JINST} \textbf{
  12} (2017) P02014,
  \href{http://dx.doi.org/10.1088/1748-0221/12/02/P02014}{\doi{10.1088/1748-0221/12/02/P02014}},
\href{http://www.arXiv.org/abs/1607.03663}{\texttt{arXiv:1607.03663}}.

\bibitem{CMS:2017wyc}
\href {https://cds.cern.ch/record/2256875}{{CMS Collaboration}, ``Jet
  algorithms performance in {13\TeV} data'',} CMS Physics Analysis Summary
  CMS-PAS-JME-16-003, 2017.

\bibitem{CMS:2019ctu}
\hrefCMSnoop {}{{CMS Collaboration}, ``Performance of missing transverse
  momentum reconstruction in proton-proton collisions at {$\sqrt{s} = 13\TeV$}
  using the {CMS} detector'',} \textit{ JINST} \textbf{ 14} (2019) P07004,
  \href{http://dx.doi.org/10.1088/1748-0221/14/07/P07004}{\doi{10.1088/1748-0221/14/07/P07004}},
\href{http://www.arXiv.org/abs/1903.06078}{\texttt{arXiv:1903.06078}}.

\bibitem{CMS:2018jrd}
\hrefCMSnoop {}{{CMS Collaboration}, ``{Performance of reconstruction and
  identification of $\tau$ leptons decaying to hadrons and $\nu_\tau$ in
  $\Pp\Pp$ collisions at $\sqrt{s}= 13\TeV$}'',} \textit{ JINST} \textbf{ 13}
  (2018) P10005,
  \href{http://dx.doi.org/10.1088/1748-0221/13/10/P10005}{\doi{10.1088/1748-0221/13/10/P10005}},
  \href{http://www.arXiv.org/abs/1809.02816}{\texttt{arXiv:1809.02816}}.

\bibitem{CMS:2022prd}
\hrefCMSnoop {}{{CMS Collaboration}, ``{Identification of hadronic tau lepton
  decays using a deep neural network}'',} \textit{ JINST} \textbf{ 17} (2022)
  P07023,
  \href{http://dx.doi.org/10.1088/1748-0221/17/07/P07023}{\doi{10.1088/1748-0221/17/07/P07023}},
  \href{http://www.arXiv.org/abs/2201.08458}{\texttt{arXiv:2201.08458}}.

\bibitem{ref:dbscan}
\hrefCMSnoop {}{M.~Ester, H.-P. Kriegel, J.~Sander, and X.~Xu, ``{A
  density-based algorithm for discovering clusters in large spatial databases
  with noise}'',} in \textit{ \it Proc. $2^{nd}$ Int. Conf. on Knowledge
  Discovery and Data Mining}, p.~226.
\newblock 1996.

\bibitem{ref:ftest}
\hrefCMSnoop {}{R.~A. Fisher, ``On the interpretation of $\chi^{2}$ from
  contingency tables, and the calculation of {P}'',} \textit{ J. R. Stat. Soc.}
  \textbf{ 85} (1922) 87,
  \href{http://dx.doi.org/10.2307/2340521}{\doi{10.2307/2340521}}.

\bibitem{Baker:1983tu}
\hrefCMSnoop {}{S.~Baker and R.~D. Cousins, ``Clarification of the use of chi
  square and likelihood functions in fits to histograms'',} \textit{ Nucl.
  Instrum. Meth.} \textbf{ 221} (1984) 437,
  \href{http://dx.doi.org/10.1016/0167-5087(84)90016-4}{\doi{10.1016/0167-5087(84)90016-4}}.

\bibitem{lindsey1996parametric}
J.~K. Lindsey, ``Parametric statistical inference''.
\newblock {Oxford Science Publications}. Clarendon Press, 1996.
\newblock ISBN~9780198523598.

\bibitem{CMS-LUM-17-003}
\hrefCMSnoop {}{{CMS Collaboration}, ``Precision luminosity measurement in
  proton-proton collisions at {$\sqrt{s} =13\TeV$} in 2015 and 2016 at
  {CMS}'',} \textit{ Eur. Phys. J. C} \textbf{ 81} (2021) 800,
  \href{http://dx.doi.org/10.1140/epjc/s10052-021-09538-2}{\doi{10.1140/epjc/s10052-021-09538-2}},
  \href{http://www.arXiv.org/abs/2104.01927}{\texttt{arXiv:2104.01927}}.

\bibitem{CMS-PAS-LUM-17-004}
\href {https://cds.cern.ch/record/2621960/}{{CMS Collaboration}, ``{CMS}
  luminosity measurement for the 2017 data-taking period at {$\sqrt{s} =
  13\TeV$}'',} CMS Physics Analysis Summary CMS-PAS-LUM-17-004, 2018.

\bibitem{CMS-PAS-LUM-18-002}
\href {https://cds.cern.ch/record/2676164/}{{CMS Collaboration}, ``{CMS}
  luminosity measurement for the 2018 data-taking period at {$\sqrt{s} =
  13\TeV$}'',} CMS Physics Analysis Summary CMS-PAS-LUM-18-002, 2019.

\bibitem{JUNK1999435}
\hrefCMSnoop {}{T.~Junk, ``{Confidence level computation for combining searches
  with small statistics}'',} \textit{ Nucl. Instrum. Meth. A} \textbf{ 434}
  (1999) 435,
  \href{http://dx.doi.org/10.1016/S0168-9002(99)00498-2}{\doi{10.1016/S0168-9002(99)00498-2}},
  \href{http://www.arXiv.org/abs/hep-ex/9902006}{\texttt{arXiv:hep-ex/9902006}}.

\bibitem{Read_2002}
\hrefCMSnoop {}{A.~L. Read, ``Presentation of search results: The {\CLs}
  technique'',} \textit{ J. Phys. G} \textbf{ 28} (2002) 2693,
\href{http://dx.doi.org/10.1088/0954-3899/28/10/313}{\doi{10.1088/0954-3899/28/10/313}}.

\bibitem{CMS-NOTE-2011-005}
\href {https://cds.cern.ch/record/1379837}{{ATLAS and CMS Collaborations, and
  LHC Higgs Combination Group}, ``Procedure for the {LHC} {Higgs} boson search
  combination in {Summer} 2011'',} Technical Report CMS-NOTE-2011-005,
  ATL-PHYS-PUB-2011-11, 2011.

\bibitem{Cowan:2010js}
\hrefCMSnoop {}{G.~Cowan, K.~Cranmer, E.~Gross, and O.~Vitells, ``{Asymptotic
  formulae for likelihood-based tests of new physics}'',} \textit{ Eur. Phys.
  J. C} \textbf{ 71} (2011) 1554,
  \href{http://dx.doi.org/10.1140/epjc/s10052-011-1554-0}{\doi{10.1140/epjc/s10052-011-1554-0}},
  \href{http://www.arXiv.org/abs/1007.1727}{\texttt{arXiv:1007.1727}}.
  [Erratum: \DOI{10.1140/epjc/s10052-013-2501-z}].

\bibitem{CMS:2024onh}
\hrefCMSnoop {}{{CMS Collaboration}, ``{The CMS statistical analysis and
  combination tool: \textsc{Combine}}'',} \textit{ Comput. Softw. Big Sci.}
  \textbf{ 8} (2024) 19,
  \href{http://dx.doi.org/10.1007/s41781-024-00121-4}{\doi{10.1007/s41781-024-00121-4}},
  \href{http://www.arXiv.org/abs/2404.06614}{\texttt{arXiv:2404.06614}}.

\bibitem{Verkerke:2003ir}
\href
  {https://www.slac.stanford.edu/econf/C0303241/proc/papers/MOLT007.PDF}{W.~Verkerke
  and D.~Kirkby, ``The \textsc{RooFit} toolkit for data modeling'',} in
  \textit{ {Proc. 13th International Conference on Computing in High Energy and
  Nuclear Physics (CHEP 2003): La Jolla CA, United States, March 24--28,
  2003}}.
\newblock 2003.
\newblock
  \href{http://www.arXiv.org/abs/physics/0306116}{\texttt{arXiv:physics/0306116}}.
\newblock [eConf C0303241 (2003) MOLT007].

\bibitem{Moneta:2010pm}
L.~Moneta\hrefCMSnoop {}{ { et~al.}, ``The \textsc{RooStats} project'',} in
  \textit{ {Proc. 13th International Workshop on Advanced Computing and
  Analysis Techniques in Physics Research (ACAT 2010): Jaipur, India, February
  22--27, 2010}}.
\newblock 2010.
\newblock \href{http://www.arXiv.org/abs/1009.1003}{\texttt{arXiv:1009.1003}}.
\newblock [PoS (ACAT2010) 057].
  \href{http://dx.doi.org/10.22323/1.093.0057}{\doi{10.22323/1.093.0057}}.

\end{thebibliography}\endgroup
\cleardoublepage \appendix\section{The CMS Collaboration \label{app:collab}}\begin{sloppypar}\hyphenpenalty=5000\widowpenalty=500\clubpenalty=5000
\cmsinstitute{Yerevan Physics Institute, Yerevan, Armenia}
{\tolerance=6000
V.~Chekhovsky, A.~Hayrapetyan, V.~Makarenko\cmsorcid{0000-0002-8406-8605}, A.~Tumasyan\cmsAuthorMark{1}\cmsorcid{0009-0000-0684-6742}
\par}
\cmsinstitute{Institut f\"{u}r Hochenergiephysik, Vienna, Austria}
{\tolerance=6000
W.~Adam\cmsorcid{0000-0001-9099-4341}, J.W.~Andrejkovic, L.~Benato\cmsorcid{0000-0001-5135-7489}, T.~Bergauer\cmsorcid{0000-0002-5786-0293}, S.~Chatterjee\cmsorcid{0000-0003-2660-0349}, K.~Damanakis\cmsorcid{0000-0001-5389-2872}, M.~Dragicevic\cmsorcid{0000-0003-1967-6783}, P.S.~Hussain\cmsorcid{0000-0002-4825-5278}, M.~Jeitler\cmsAuthorMark{2}\cmsorcid{0000-0002-5141-9560}, N.~Krammer\cmsorcid{0000-0002-0548-0985}, A.~Li\cmsorcid{0000-0002-4547-116X}, D.~Liko\cmsorcid{0000-0002-3380-473X}, I.~Mikulec\cmsorcid{0000-0003-0385-2746}, J.~Schieck\cmsAuthorMark{2}\cmsorcid{0000-0002-1058-8093}, R.~Sch\"{o}fbeck\cmsAuthorMark{2}\cmsorcid{0000-0002-2332-8784}, D.~Schwarz\cmsorcid{0000-0002-3821-7331}, M.~Sonawane\cmsorcid{0000-0003-0510-7010}, W.~Waltenberger\cmsorcid{0000-0002-6215-7228}, C.-E.~Wulz\cmsAuthorMark{2}\cmsorcid{0000-0001-9226-5812}
% ...
\par}
\vskip\cmsinstskip
\dag:~Deceased\\
$^{1}$Also at Yerevan State University, Yerevan, Armenia\\
$^{2}$Also at TU Wien, Vienna, Austria\\
% ...
\end{sloppypar}
\end{document}